\documentclass[12pt,preprint]{aastex}
  \usepackage{rotating}
  \usepackage{graphicx}
  \usepackage{dcolumn}
  \usepackage{natbib}

  \usepackage{natbib}
  \bibpunct{(}{)}{;}{a}{}{,} 

\newcommand{\fluxunitsAA}{10$^{-16}$ erg~s$^{-1}$~cm$^{-2}$}
\newcommand{\kms}{km s$^{-1}$}
\newcommand{\Hab}{H$\alpha$/H$\beta$}
\newcommand{\AB}{H$\alpha$/H$\beta$}
\newcommand{\OB}{[OIII]$\lambda$5007/H$\beta$}
\newcommand{\NA}{[NII]$\lambda$6583/H$\alpha$}
\newcommand{\Srat}{[SII]$\lambda$6716/[SII]$\lambda$6731}

\begin{document}

\title{Integral Field Spectroscopy of the central regions of 3C~120: Evidence of a past merging event}

\author{B.Garcia-Lorenzo\altaffilmark{1}, S.F~S\'anchez\altaffilmark{2},  
 E.Mediavilla\altaffilmark{1}, J.I. Gonz\'alez-Serrano\altaffilmark{3,4}, 
L.Christensen\altaffilmark{2}}

\altaffiltext{1}{Instituto de Astrof\'\i sica de Canarias, 38205 La Laguna,
  Tenerife, Spain}
\email{bgarcia@ll.iac.es}
\altaffiltext{2}{Astrophysikalisches Institut Potsdam, An der Sternwarte 16, 14482 Potsdam, Germany}
\altaffiltext{3}{Instituto de Fisica de Cantabria, UC-CSIC, Av. de Los Castros
  S/N, 35005, Santander, Spain}
\altaffiltext{4}{Dept. de Fisica Moderna, Facultad de Ciencias, UC, Av. de Los Castros
  S/N, 35005, Santander, Spain}

\begin{abstract}
  
  Optical integral field spectroscopy (IFS) combined with Hubble Space
  Telescope (HST) WFPC imaging were used to characterize the central regions of
  the Seyfert 1 radio galaxy 3C~120. We carried out the analysis of the data,
  deriving intensity maps of different emission lines and the continua
  at different wavelengths from the observed spectra. Applying a 2D modeling
  to the HST images we decoupled the nucleus and the host galaxy, and analyzed
  the host morphology. The host is a highly distorted bulge dominated galaxy,
  rich in substructures. We developed a new technique to model the IFS data
  extending the 2D modeling (3D modeling hereafter). Using this technique we
  separated the Seyfert nucleus and the host galaxy spectra, and derived a
  residual data cube with spectral and spatial information of the different
  structures in 3C~120. Three continuum-dominated structures (named A, B, and
  C) and other three extended emission line regions (EELRs, named E$_1$, E$_2$
  and E$_3$) are found in 3C~120 which does not follow the general behavior of
  a bulge dominated galaxy. We also found shells in the central kpc that may
  be remnants of a past merging event in this galaxy. The origin of E$_1$ is
  most probably due to the interaction of the radio-jet of 3C~120 with the
  intergalactic medium \citep{axon89,sanc04c}. Structures A, B, and the shell
  at the southeast of the nucleus seem to correspond to a larger
  morphological clumpy structure that may be a tidal tail, consequence of the
  past merging event. We found a bright EELR (E$_2$) in the innermost part of
  this tidal tail, nearby the nucleus, which shows a high ionization level.
  The kinematics of the E$_2$ region and its connection to the tidal tail
  suggest that the tail has channeled gas from the outer regions to the
  center.

\end{abstract}

\keywords{galaxies: individual: 3C~120 --- galaxies: ionized gas and stellar kinematics and 
dynamics --- galaxies: Seyfert }

\section{Introduction}

Radio galaxies are excellent laboratories to disentangle the role of central
active nuclei in host galaxies evolution and the relation of intergalactic
environment and the central activity. The complex structure of radio galaxies
put severe troubles to carry out specific optical studies in this matter.
Fortunately, the recent technical and computational improvements on integral field spectroscopy (IFS hereafter) open new frontiers to multitude of astronomical projects.
In the particular case of radio galaxies, IFS data allow to separate the
different components and to perform an independent analysis of the various
subsystems and their evolution.

3C~120 is a Seyfert 1 galaxy at a redshift of $z=0.033$. It has been
 studied extensively at many wide-range wavelengths, with special attention in
X-ray and radio frequencies \citep[e.g.][]{walk87,mole88,walk97,ogle04}.
Although 3C~120 was classified as an early-type galaxy based on a
visual inspection of ground-based images \citep{zwic71}, \cite{sarg67}
reported a faint spiral structure.  However, its optical morphology is still
not clear showing an elliptical shape with some peculiarities \citep{hansen87}
and could be a merger remnant galaxy \citep{mole88}.  Spectroscopic analysis
has shown that 3C~120 presents a rotation curve most likely indicating the
presence of an undetected disk \citep{mole88}. Based on this result, 3C~120 is
normally quoted as an early-type spiral or S0 galaxy \citep{mole88}.

Although Seyfert galaxies are generally considered as radio-quiet objects,
3C~120 is an active radio-loud source showing a superluminal
one-sided jet that extends $\sim$25$\arcsec$ out of the core
\citep{read79,walk88,walk97}. 
Different authors \cite[e.g.][]{bald80,pere86,soub89,hua88} reported the
existence of several continuum and emission line dominated structures in this
object. Some of these structures have a clear relation with the radio-jet.
\cite{hjor95} detected a continuum dominated optical counterpart of the
radio-jet using deep broad-band ground based images. On the other hand, an
EELR (E$_1$) associated with the jet was found at $\sim$5$\arcsec$ west of
the nucleus \citep{hua88,axon89,sanc04c}.  The nature of the remaining
structures is still unclear \citep[e.g.][]{pere86,mole88}.

This paper compiles data from IFS optical observations of 3C~120 using the
optical fibers system INTEGRAL covering a central region of
$\sim16\arcsec\times12\arcsec$ field-of-view. We combined our IFS data with
high-resolution Hubble Space Telescope (HST) images. We describe the optical
fibers instrument, observations, and data reduction in section \S2. In section
\S3, we describe the emission line profiles in the circumnuclear region of
3C~120 and we present 2D spectra diagrams for different emission lines
(appendix A). In the following sections, we present the stellar morphology and
the structures detected (section \S4.1), the average colors of the galaxy
and the colors of those structures (section \S4.2), and the emission-line
morphology and ionization structure (section \S4.3). In section \S5, we apply
a new technique to separate the spectra of different components coexisting in
3C~120 and analyze the properties of each component. We present the velocity
field of the central region of 3C~120 and the discussion of several
kinematic perturbations in section \S6. A distance of 198 Mpc is assumed for
3C~120 throughout the paper (H$_0$=50 km s$^{-1}$ Mpc$^{-1}$) which
corresponds to a scale of $\sim 1 $kpc/$\arcsec$.

\section{Experimental Set-Up, Observations, and Data Reductions}

\subsection{Integral Field Spectroscopy}

IFS optical observations with fibers are based on the idea of
connecting the focal plane of the telescope with the spectrograph slit using a
fiber bundle. In this way, when an extended object is observed, each fiber
receives light coming from a particular region of the object. Each individual
spectrum appears well separated on the detector and therefore, spatial and
spectral information are collected simultaneously. The wavelength limitations
essentially depend on the characteristics of the spectrograph itself. The
spatial resolution depends on the fiber sizes and the prevailing seeing
conditions during the observations. The spatial coverage of fiber systems is
relatively small \citep[see ][ for a counter example]{lefe03}, but they are
very useful when studying small size objects, such us the circumnuclear region
of nearby active galaxies \citep[e.g.][]{garcia01}, blue compact dwarf
galaxies \citep[e.g.][]{cairos02} or gravitational lenses
\citep[e.g.][]{motta02,wiso03}.

IFS of 3C~120 was obtained the 26th of February 2003 at the Observatorio del
Roque de los Muchachos (ORM) on the island of La Palma, Spain. The 4.2m
William Herschel Telescope (WHT) was used in combination with the fiber system
INTEGRAL \citep{arr98,arr99,medi98} and the WYFFOS spectrograph
\citep{Bing94}. The observations were carried out under photometric conditions
and an average seeing of 1$\farcs$2. The standard bundle 2 of INTEGRAL was
used during these observations. This bundle consists of 219 fibers, each one
with a diameter of 0$\farcs$9 projected on the sky. A central rectangle is
formed by 189 fibers covering a field-of-view of $16\arcsec\times12\arcsec.3$,
and the remaining 30 fibers form a ring with a diameter of $90\arcsec$. Figure
1 illustrates the actual distribution of the science fibers on the
focal plane.

The WYFFOS spectrograph was equipped with a 300 groove mm$^{-1}$ grating
centred on 5500 \AA\ (spectral coverage: 3500-9000 \AA ). A Tek6 CCD array of
$1124\times1124$ pixels of 24 $\mu$m size was used, giving a linear dispersion
of about 3 \AA \ pixels$^{-1}$.  With this configuration, and pointing to the
central region of 3C~120, three exposures of 1800 seconds each were taken.
The data were reduced using IRAF standard routines \citep{tody86}. Although
the reduction of IFS data from fiber-based instruments does not differ
significantly from standard spectroscopic data reduction, we describe briefly
in this section the reduction procedure.

A master bias frame was built by averaging different bias frames taken along
the night. This bias frame was then subtracted from the science frames. In
observations with optical fibers, flats-fields are obtained illuminating the
focal plane uniformly, and obtaining spectra (the so-called flat-spectra).
Thus, for a particular wavelength, the differences in response among fibers
dominate the flat-spectra shape. These differences are due to their distinct
focal-ratio degradation, position at the entrance of the spectrograph, etc. In
this way, flat-spectra are used to homogenize the response of all the fibers.
Flat-spectra are also used to obtain the polynomial fits to define the fiber
path along the detector and extract the individual spectra from the whole
image. Each spectrum appears well separated on the detector, with a width of
approximately two pixels in the spatial direction (according to the core fiber
image size of the fiber bundle used for these observations). The trace and
extraction of individual spectra was performed using the standard routine
APALL of IRAF. After this operation, frames of $1124\times1124$ pixels were
reduced to $1040\times219$ pixel: each pixel in the spatial direction contains
the spectrum of each particular fiber, a total of 219 fibers.
 
The wavelength calibration was done using the IDENTIFY and REIDENTIFY routines
of IRAF.  In order to carry out this procedure, we selected several isolated
and well-distributed arc lines. Although it might be difficult to determine
the actual uncertainty produced by the wavelength calibration, we have used
the sky lines in our spectra to estimate the final wavelength errors, being
smaller that 35.5 kms$^{-1}$ and 29.5 kms$^{-1}$ at blue and red wavelengths,
respectively.

The SP1045+378 flux standard star (Isaac Newton Group Database) was observed
the same night and under similar conditions as for 3C~120.  The standard star
has been used to calibrate a flux ratio by comparing with the standard-star
flux tables of Stone (1977). We reduced the SP1045+378 frames in a similar way
than the object ones, correcting them for differential atmospheric refraction
\citep{fili82} using E3D \citep{sanc04}. Then, we extracted the observed
spectrum of the star by co-adding the spectra of the central 37 fibers, which
includes $>$99\% of the total flux. Comparing this observed spectrum with the
flux-calibrated spectrum of SP1045+378 we derived a sensitivity curve that we
used to calibrate the object frames.

We combine the IDA tool \citep{garcia02} and the Euro3D visualization package
\citep{sanc04} to analyze the data and generate two-dimensional maps of any
spectral feature (intensity, velocity, width, etc). Maps recovered from
spectra are images of 51$\times$37 pixels with a scale 0$\farcs$3/pixel. While
the spatial sampling of the used INTEGRAL configuration is 0$\farcs$9 (that
is, the fiber diameter of the bundle 2), the centroid of any peak in our maps
can be measured with an accuracy of around 1/5 of the fiber diameter, that is
$\sim 0\farcs 2$ \citep[e.g.][]{medi98}.

\subsection{HST imaging}

Wide-Field Planetary (WFPC) camera images of 3C~120 in different bands are
available on the archive of the Hubble Space Telescope (HST). We obtained
these images to study the morphology of this object. Table \ref{tab_hst}
summarizes the properties of these images. The data set comprises three
broadband images (F555W, F675W and F814W), which roughly corresponds to the
standard $V$, $R$ and $I$-bands, and a medium band image (F547M). These images
basically sample the continuum emission in 3C~120, since the equivalent width
of mostly all the emission lines is very small compared with the width of the
bands. However, there is a non neglectible contamination, dominated by the
[OIII] and H$\beta$ emission lines in the F555W-band image (and H$\alpha$ in
the F657W-band image). The F547M-band image can be used as an estimation of
the pure continuum emission, since the emission lines are at the edge of its
transmission curve. The data set is deep enough to study the different
structures present in this object \citep{soub89}.

\section{Atlas of Spectra}

Figure 2 shows the nuclear spectrum of 3C~120 in the full
wavelength range. This spectrum has been obtained by co adding the seven
spectra closest to the continuum peak. This is almost equivalent to a hexagonal
aperture of 1$\farcs$6 in radius centred at the optical nucleus of 3C~120.
We can easily recognize several emission lines and the characteristic broad
component of permitted emission lines of Seyfert 1 galaxies.

In the Appendix A, we present the individual spectra corresponding to each of
the different observed positions (fibers) in selected spectral intervals that
include the most important emission lines (spectra diagrams).

\subsection{Emission-Line Profiles}

The profiles of the emission lines in the nuclear spectrum of 3C~120 show a
considerable blending due to the wings of the broad component of permitted
lines (Fig. 2). The H$\delta$ profile appears to be
considerably broader than any other Balmer line, but this is not the case of
H$\delta$ as in previously found by other studies \citep{phil75,bald80}. The
HeI$\lambda5876$ shows also a considerable broadening.  At this step, we cannot
discuss the origin of these features in terms of recombination models and
reddening effects because the nuclear spectrum in figure
2 includes the contribution of the nucleus and the
surrounding galaxy. In section \S5 we separate both contributions and we will
be in position to tackle this discussion. The low spectral resolution of the
current IFS data prevents the detection of double peaks reported by
\cite{axon89}. However, the emission lines show asymmetric profiles in some
locations outside the central region.

\section{Data analysis and results}

\subsection{Broad band distribution and morphological structures.}
\label{bb_hst}

Broad-band images of 3C~120 were recovered from our IFS data by coadding the
flux in spectral ranges that mimic the band-passes of the previously refereed
HST images and using an interpolation routine \citep{sanc04,garcia02}. The
specific wavelength ranges for each band were 3900-4900 \AA ($B$-band),
5000-6000 \AA ($V$-band), 5600-5700 \AA ($V'$-band), 6100-7100 \AA ($R$-band)
and 7550-8550 \AA ($I$-band). All images show a bright nucleus on top of
a weak host galaxy. Figure 3A presents the intensity map
corresponding to the $V$-band filter.  The intensity contours have an
elliptical shape except for those at around 4 arcsec from the central peak,
showing an elongation toward the west.  There are remarkable similarities
between the restored map (Fig.  3A) and the F555W-band image
from the HST (Fig.  3B), despite of the differences in
wavelength ranges and spatial sampling. The superior resolution of the HST
images allows to directly detecting several structures. These structures have been previously reported
subtracting a galaxy template to broad and narrow band ground base images
\citep[e.g.][]{bald80,pere86,hua88,soub89}. We have labeled, using the
nomenclature introduced by \cite{soub89}, the different detected structures
(figure 3B). The HST/F555W-band image allows to resolve them,
showing a richer level of structures. In particular, structures B, and C
are composed of a smoothed low surface brightness component and more luminous
clumpy substructures. Structure A seems to be more collimated, as previously
noticed by ground based imaging, which explains why it was initially confused
with the optical counterpart of the radio jet \citep[e.g.][]{pere86}. It shows
also a low surface brightness component and four clear clumps. We overplot in
figure 3B the radio-map at 4885 MHz (Walker 1997), showing that
the structure A is most likely related to the structure B rather than to the
radio-jet.  Indeed, the HST images suggest that both structures belong to a
sequence of clumpy knots which are physically connected. Dust lines and shell structures
not detected from the ground are seen in the inner regions. Two shell arcs at
$\sim1\arcsec$ north and south of the central peak are clearly detected in the HST
images. Although previous HST images show extensions to the north-west and south-east in the inner region of 3C~120 \citep{zir98}, to our knowledge this is the first time that the shell structure is clearly visible. The north and south shells are named S$_N$ and S$_S$, respectively, hereafter.  The S$_S$ seems to be connected to structure A by a faint tail.

A 2D modeling of the broad-band images was performed using GALFIT
\citep{peng02} to obtain a clear picture of the morphology of 3C~120. This
program has been extensively tested in the image decomposition of QSO/hosts
\citep{sanc04b}.  The 2D model comprises a narrow Gaussian function (to model
the nucleus) and a galaxy template (to model the galaxy) both convolved with a
PSF. We performed the fit twice, first using a Sersic law \citep{sers68} to
characterize the galaxy and then using a de~Vaucouleurs law \citep{deva48}.
This method allows us to determine the morphological type of the host galaxy,
based on the obtained Sersic index, and to get a good determination of the
galaxy flux, based on the modeling the de~Vaucouleurs law \citep{sanc04b}.
The PSF was created using a field star for the analysis of the HST images, and a
calibration star for the analysis of the broad-band images recovered from the
IFS data.

We applied first the fitting technique to the HST images because of their
better spatial resolution. 
The nucleus was saturated in the HST/F814W-band, which prevents us to do a
two-model fitting, and limits the reliability of any morphological
classification, which strongly depends on the shape of the profile in the
inner regions. To derive a rough estimation of the host magnitude we masked
the nucleus, and fixed the scale to the average of the values derived in the
other available HST/bands.  Further checks, explained below, demonstrated that
this approach was valid.  Table \ref{tab_galfit} summarizes the results from
this analysis. For each band it shows the derived Sersic index, the nucleus
and host magnitudes and the effective radius of the host galaxy.  The derived
Sersic indices, all near or larger than the nominal value of 4 for an
early-type galaxy, confirm the morphological classification of the host galaxy
of 3C~120 as a bulge dominated galaxy.

We subtracted the object template (galaxy+nucleus) derived by the 2D modeling
from the original images, obtaining a residual image for each band. These
residual images were used to study the properties of the different structures,
once decontaminated from the smooth component. As an example, the residual
image of the $V$-band and the HST/F555W-band are shown in Figure
3C and 3D, respectively. The already quoted
structures are now clearly identified.  The S$_S$ shell coincides with an
extended emission line region (EELR) previously detected in this object
\citep{hua88,soub89,sanc04c} and labeled as E$_2$ \citep{soub89}.

\subsection{Colors and gaseous distributions in 3C~120}

As already quoted all the broad-band images recovered from the IFS data show a
similar morphology than the $V$-band image shown in figure 3A.
Combining these recovered broad-band maps, we obtained color maps of 3C~120.
Figure 4A shows the $V-I$ color map derived from the IFS data. It
shows a blue circumnuclear region elongated to the southwest. The elongation
towards the west seen in the broad band images has also a counterpart in the
color maps showing bluer colors at a region $\sim4\farcs5$ west of the nucleus
than its surroundings. Figure 4B presents the $V-I$ color map
derived from the F555W and F814W-band HST images.  Despite of their different
spatial resolution, figures 4A and 4B present remarkable
similarities.  The different structures quoted in section \S4.1 show bluer
$V-I$ colors than the average color of the host galaxy, as already noticed
\citep[e.g.][]{fraix91}.  Figure 4B
also shows evidence for two weaker shells farther from the central peak than S$_N$
and S$_S$ but also at the north and south. However, these ``secondary'' shells
are fainter, being at the detection limit.

Both the $V$-band and the F555W-band images are contaminated by the emission
from [OIII] and H$\beta$. A rough estimation of how strong that
contamination is can be obtained by subtracting a scaled continuum image, clean
of those contaminations. Figure 4C shows the residual image of the
INTEGRAL $V$-band image after the subtraction of an adequate continuum (the
$V'$-band image). A similar estimation was done for the HST images,
subtracting the continuum dominated F547M-band image from the F555W-band image
(Figure 4D). This image was smoothed using an 11$\times$11 pixels
median kernel to increase the signal-to-noise ratio. The resulting maps are a
rough estimation of the distribution of the [OIII]+H$\beta$ emission. The
position of two of the EELRs detected in this object (E$_1$ and E$_2$,
hereafter) is indicated \citep{soub89,sanc04c}. The residual images of the
continuum, after subtracting a model template (Fig. 3C and
3D), are overplotted for comparison purposes. The S$_N$ and
S$_S$ shells show a strong gaseous emission in the HST images, which in the
S$_S$ extends towards the E$_2$ region. 

Figure 5 shows the $V-R$ colors of the host galaxy and the
different structures as a function of the $R-I$ colors. As already noticed
in the color image, the structures have bluer $V-R$ colors than the host
galaxy. This indicates most probably the presence of somewhat younger stellar
populations associated with them. However, they show a wider range of $R-I$
colors than that of $V-R$ colors. Indeed, the average $R-I$ color of the host
galaxy is within the range of $R-I$ colors of the different structures.  The
color-to-color distribution derived from synthetic models has been included in
the figure for comparisons. The discontinuous line shows the colors of single
stellar populations calculated using the \cite{bruz03} models. We assumed a
solar metalicity and a \cite{chab03} IMF.  The labels indicate the logarithm
of the stellar population ages in Gyrs. In general terms, the color-to-color
distribution does not match with that simple model, apart from the case of the
S$_N$ and S$_S$ shells. For the remaining structures, the $R-I$ colors
correspond on average to a young stellar population of $\sim$6$\times$10$^8$
Gyr, but the $V-R$ colors correspond to older stellar populations. This may
indicate that the real populations are a composite of populations. The dotted
line shows the colors of different galaxy types (E, Sab, Sbc, Scd and Irr)
\citep{fuku95}, which mainly correspond to different mixes of stellar
populations.  The $R-I$ color of the host galaxy mainly corresponds to a
Sab-Scd galaxy. On average, the structures have $V-R$ colors that correspond
to the same kind of galaxies (Sab-Scd), but their $R-I$ colors expand over all
the range of possible colors.

So far we have ignored the effects of dust in the color-to-color distribution
shown in Figure 5. However, the dust content in 3C~120 is
rather high; an average value of A$_V\sim 4$ mags has been estimated
\citep{sanc04c}. Dust extinction redden both $V-R$ and $R-I$ colors in an
almost similar way \citep{fitz99}. The effects of dust in the color-to-color
distribution is illustrated with an arrow in Fig. 5. It is
clear that the dust is not homogeneously distributed in the galaxy: e.g.,
typical dust lanes are seen in figure 3B. A combined effect of
different stellar populations and a non-homogeneous distribution of dust could
explain the observed color-to-color distribution. In that case, the S$_N$ and
S$_S$ structures would be dust-free areas dominated by a single stellar
population.

However, results from broadband colors should be taken with care. The
contamination of broadband filters by strong emission lines can drastically
affect the morphology of color maps. In the case of 3C~120 the problem is even
worse because of the contribution and contamination from the broad component
of the Balmer emission lines in the central regions.

\subsection{Line-Intensity Maps and Ionization Structure}

We performed a line profile fitting and de-blending in order to study the
integrated high and low ionization gas distribution, their physical properties
and kinematics.  We fit a single Gaussian to any of the bright emission lines
in the spectra.  For the Balmer lines, we included a second broader Gaussian
to fit the wide line from the broad line region (BLR). In spite of our poor
spectral resolution, we found that several spectra have evidence of
substructures, showing asymmetric profiles with blue or red wings indicating
the presence of several gaseous systems in 3C~120. Indeed, previous authors
(Axon et al.  1989) reported the existence of several components in the emission lines of 3C~120, but to apply a line-profile decomposition to separate the
different gaseous components would be unrealistic because of the low spectral
resolution of the current IFS data. The broad component of the Balmer lines is
confined to the nucleus discarding an extended broad emission line region.

Figure 6 shows the intensity maps of [OIII]$\lambda5007$ (figure
6A) and H$\alpha$ (figure 6B) after the
deblending. Intensity contours of ionized gas are clearly elongated toward the
E$_2$ structure. Emission line maps present an elongation to the west, forming
a clear secondary peak at around 5 arscec from the nucleus in the [OIII]
intensity map. This secondary peak corresponds to the E$_1$ structure
previously quoted (section \S4.2). The ratio of the H$\alpha$ intensity of the
nucleus and E$_1$ is much smaller than that of [OIII]$\lambda5007$, suggesting
a high ionization nature of the latter. The nuclear region shows a low
ionization degree in the [OIII]$\lambda5007$/H$\beta$ map surrounded by a ring
of high ionization (figure 6C), with larger values at the south
of the nucleus, coincident with the location of E$_2$. A high ionisation
region expands from the ring to the west, increasing the ionisation degree
along this direction. The [NII]$\lambda6584$/H$\alpha$ line ratio map (figure
6D) also shows a ring structure, surrounding the nuclear
region. The E$_1$ is on the path of the 3C~120 radio-jet, as well as the
continuum dominated structure {\it A}. 

The high [OIII]/H$\beta$ ratio of E$_1$ (figure 6C) discards a
relation between {\it A} and E$_1$, assuming that {\it A} is a star forming
region in a spiral arm or a tidal tail, and points, more likely, to a direct
connection between E$_1$ and the radio-jet \citep{sanc04b}.  While the nucleus
and the {\it A} region present [OIII]$\lambda5007$/H$\beta$ and
[NII]$\lambda6584$/H$\alpha$ ratios at the limit of HII regions in a
diagnostic diagram \citep{veil87}, those ratios of E$_1$ and its west surroundings are placed on the high-ionization region .

The distribution of the H$\alpha$/H$\beta$ line ratio (figure
6E) shows a dust lane crossing 3C~120 along the
southeast-northwest direction. The high Balmer decrement at the west points to
a high dust obscuration around E$_1$.  However, we cannot rule out that the
high H$\alpha$/H$\beta$ ratio may be related to the low signal to noise of the
narrow component of H$\beta$ in the circumnuclear region of the galaxy. The
electronic density derived from the [SII]$\lambda6716$/$\lambda6730$ presents
a patchy structure, showing an enlargement in the location of E$_1$. In the next
section, the ionization conditions of the different structures in the observed
region of 3C~120 are studied in detail.

\section{Decoupled spectra of the different components in 3C~120}

In the previous section, we used the traditional method (Gaussian fitting) to
deblend the different gaseous components. IFS optical observations record
spatial and spectral information simultaneously. Taking advantage of this
fact, we have developed a technique (3D-modeling hereafter) to disentangle
the spectra of the main components of 3C~120 (nucleus+host), and to obtain
cleaned spectra of the different structures described above. This technique
has been successfully applied recently \citep{sanc04c}. A brief summary of the
technique is described in the Appendix B, and will be explained in
detail in a separate article (S\'anchez et al., in prep.).  In summary, the
technique provides us with a spectrum of each of the main components of the
object plus a residual data cube that can be used to derive cleaned spectra of
the different structures in the object and/or analyze their morphological and
kinematical properties.  This is an extension of the 2D modeling of images
applied in section \S4.1 to each of the monochromatic images of the data cube
derived from IFS optical observations. We applied the 3D modeling to the
current data and we derived the 3C~120 nucleus spectrum, the mean host galaxy
spectrum and a residual data-cube of spectra.

\subsection{Morphological structures from the residual data cube of spectra}

To compare the results from the proposed technique to those of the traditional
method, figure 7A shows the greyscale of the residual from the 2D
modeling of the F555W/HST-band image obtained in section \S4.1 (figure
3D). We over plotted the contours of a narrow-band image
centred on the continuum adjacent to the [OIII] emission line (5204-5246 \AA)
extracted from the residual data cube of spectra obtained after the 3D
modeling. The main continuum dominated structures (A, B and C) are
detected $\sim$1$\arcsec$ from the nucleus.  As expected, the shell
structure detected at $\sim$1$\arcsec$ north and south of the nucleus in the
HST images is not detected with the IFS data. A combination of the superior
resolution of the HST images, the effects of the seeing and the sampling of
our IFS data can explain it. In figure 7B we present the
greyscale of the F555W$-$F547M image (obtained in section \S4.2, figure
4D) and the contours of a narrow-band image centred on the
[OIII]$\lambda$5007 emission line (5170-5200\AA) extracted from the residual
data cube. As we quoted above (section \S4.2), the F555W-F547M image is a
rough estimation of the [OIII]-H$\beta$ emission map.  Although both images
are in good agreement, integral field spectroscopic data are more adequate to
detect extended emission line regions (EELRs) that are diluted in broadband
filter images, like the F555W-band one. It is possible to identify the E$_1$
and E$_2$ structures, and a third region (E$_3$ hereafter) at the north-west
of the F555W-F547M image. \cite{soub89} already detected these emission line
regions, using narrow-band imaging centred on the [OIII] emission line
spectral region. E$_1$ and E$_2$ are completely coincident with their reported
positions. The peak of E$_3$ found by \cite{soub89} is not included within the field of view of INTEGRAL. We are confident of our detection, and most probably we are seeing a tail
previously not detected of E$_3$ towards the nucleus.

\subsection{Spectra of the nucleus and the host galaxy}
\label{hg_spec}

Figure 8 and 9 show the spectra of the different
deblended components of 3C~120. Figure 8A shows the spectrum of
the nucleus. Its continuum follows roughly a power-law with a slope of
$\alpha\sim-0.5$ ($F_\lambda\propto\lambda^\alpha$). It contains
broad-emission lines, without any trace of them in the spectrum of the host
galaxy (figure 8B), which indicates that the decoupling
technique has worked properly. The intensities of the narrow emission lines is
weaker in the nucleus than in the host galaxy spectrum. Indeed, many emission
lines are clearly detected in the host galaxy spectrum but not in the nucleus.

To derive the properties of the emission lines in the spectrum of the nucleus,
we fit Gaussian functions to the different emission lines. The adjacent
continuum was modelled with a low-degree polynomial function.  To increase the
accuracy of our model, different nearby emission lines were fitted together,
defining line systems with the same systemic velocity and
full-width-half-maximum (FWHM). After several attempts, the best (and
simplest) modeling was obtained including two decoupled broad systems, with
$\sim$8796 \kms\ and $\sim$2188 \kms, and a narrow system, with $\sim$700 \kms
.  Recently, \cite{ogle04} also reported two broad and one narrow component
for the H-Balmer lines in the nucleus of this galaxy with a similar velocity
dispersions than those derived here.  Table \ref{elines1} lists the result
from the fit, including the wavelength, the flux  and their 1$\sigma$ errors
for each detected line. The uncertainties in the determination of the flux of
the narrow emission lines affected by line blending are high as expected. The
H$\alpha$/H$\beta$ line ratios are in the range of $\sim$3.0-3.6 for the
different line systems, in agreement to the nominal values for case B
recombination \citep{oste89}.  Therefore both the narrow and broad emission
line regions are not significantly affected by dust, as expected for a type 1
AGN nucleus.  We estimated an effective temperature of $\sim$2.3$\times$10$^5$
K using the ([OIII]$\lambda$5007+[OIII]$\lambda$4959)/[OIII]$\lambda$4363 line
ratio and the relation between this ratio and temperature \citep{oste89}
typical of this kind of AGN.

Figure 8B shows the average spectrum of the host galaxy together
with the spectra of different synthetic models consisting on single stellar
populations of different ages. We used the \cite{bruz03} models, assuming a
solar metallicity and a \cite{chab03} initial mass function (IMF). The effects
of dust were included in the model using the extinction curve of \cite{fitz99}
for a dust absorption of $A_V\sim$4 mags (the average value in the host
galaxy, as we will show below). Two extreme cases were plotted, representing
an old ($\sim$16Gyr, red line) and a young stellar population ($\sim$0.01Gyr,
blue line). These extreme cases are a clear over simplification, but they can
be used just for qualitative comparisons. A single stellar population cannot
describe the observed spectrum: the optical slope between $\lambda\sim$5000
\AA \ and $\lambda\sim$6500 \AA \ matches well with that of an old stellar
population.  However, the slope at larger wavelengths is flatter, more similar
to that of a young stellar population. We already noticed it (section \S4.2)
when we analyzed the broadband colors of the host galaxy (Figure
5). Only a mix of different stellar populations can explain the
observed spectrum.

We included in figure 8B the spectrum of a fifty-to-fifty mix of
both components (orange line). Even this simple model can describe quite well
the spectrum of the host galaxy for $\lambda>$4500\AA. At shorten wavelengths
there is an increase of the flux, which does not match with the mix model
(dashed line). It is well known that powerful radio galaxies show a
significant blue-UV excess \citep{lill84} most probably due to scattered light
from the nucleus than to young stellar populations \citep{mcca87}. This
scattered light can be well described by a power-law. Adding a power-law to the
mix model we can describe the average host galaxy spectrum for any wavelength
(orange solid line). Therefore, three components are needed to explain
qualitatively the spectrum of the host galaxy: (a) an underlying old stellar
population, (b) a young stellar population, and (c) a nebular continuum, most
probably scattered light from the nucleus. We did not perform a proper fitting
to determine the best fractions of the different components because it would
not change the qualitative statement, and, in any case, the specific fractions
will depend on the assumed synthetic models for the young and old populations
and we chose them arbitrary. The contribution of each component to the total
flux are $\sim$5\% (a), $\sim$55\% (b) and $\sim$40\% (c) at 4500 \AA \ and
$\sim$35\%(a), $\sim$64\% (b) and $\sim$2\% (c) at 6000 \AA . Similar results
have been found in the study of other powerful radio galaxies
\cite[e.g.][]{tadh96}.

Table \ref{elines1} lists the results derived from fitting a single Gaussian
to the emission lines in the host galaxy spectrum. As we quoted above, only
narrow emission lines corresponding to a dispersion velocity of FWHM$\sim$700
\kms \ are detected. These emission lines are considerably stronger than the
narrow-emission line region in the spectrum of the nucleus, indicating the
presence of an extended narrow-emission line region which extends through most
of the field of view. Similar results have been found in spectroscopic studies
of radio-quite type 1 AGNs \citep[e.g.][]{jahn02,jahn04}. The \Hab\ line ratio
is $\sim$10, indicating a dust absorption of $A_V\sim$4 mags when comparing
with the nominal case B recombination value \citep{oste89} and using the
\cite{fitz99} extinction curve. In average the host galaxy contains a
 rather high dust content, but the dust distribution is not uniform as we can
 see in figure 6E. We estimated an effective temperature of
$\sim$14000 K using the
([OIII]$\lambda$5007+[OIII]$\lambda$4959)/[OIII]$\lambda$4363 line ratio and
an electron density of n$_e$$\sim$10 cm$^{-3}$ using the
[SII]$\lambda$6716/6731 line ratio. To derived these quantities, we used the
relations between these line ratios and the measured parameters
\citep{oste89}. The line ratios log(\OB)$\sim$1.2 and log(\NA)=$-$0.4 indicate
most probably a direct photo ionization by the UV flux from the AGN,
consistent with the result shown in Section \S4.3.

\subsection{Spectra of the EELRs: Origin of the ionization}
\label{EELRs_spec}

The integrated individual spectra of the structures E$_1$, E$_2$, E$_3$
described in section \S5.1 were extracted from the residual data cube.  Figure
9A presents the spectra of each EELRs (E$_1$, E$_2$, E$_3$),
which are remarkable similar.  They have a clear gaseous nature, without
appreciable continuum emission. This result confirms that the continuum
dominated structure $A$ is not related to E$_1$ as we pointed out in section
\S4.2. A visual inspection of the relative strength of the different lines
indicates that the dominant ionization source is the AGN or shocks, rather
than a star formation process. This was already noted in section \S4.3 and
figure 6C for E$_1$, and E$_2$. Table \ref{elines2} lists the
result of the Gaussian modeling of the emission lines in these spectra.
Different line ratios and parameters derived from them are listed in Table
\ref{param}. We included the \OB, \NA, \Srat \ line ratios, and the derived
dust absorption ($A_V$), electron density ($n_e$) and effective temperature
(T$_{eff}$). Those values of the nucleus and the host galaxy spectra were also
included for comparison purposes. As we already mentioned, the ionization
conditions are similar in the different EELRs, being also similar to those of
the nucleus and the host. Figure 10 shows the classical diagnostic
diagram of the \OB \ line ratio as a function of \NA \ for the different
components of 3C~120, including the division between AGNs and star forming
regions \citep{veil87}. The major difference is found in the position of the
nucleus in the diagram, which seems to lie in the location of star forming
regions. This is most probably due to the large uncertainties in the \NA\ line
ratio derived for the spectrum of the nucleus, largely influenced by the
deblending process of the narrow and broad emission lines, as we quoted above.
The derived effective temperature (T$\sim1.4\times10^{4}$ K) for the host galaxy spectrum does not exclude the hot stars as a possible origin for the ionization but the high [SII]/H$\alpha$, [NII]/H$\alpha$, and [OIII]/H$\beta$ ratios 
indicate that the average ionization for the host is the AGN or/and a shock
 process. i.e., the observed spectra cannot be found in HII regions.
The \NA \ line ratio of the E$_3$ region is lower than that value for the rest
of the structures. E$_3$ is the faintest detected EELR and it was not
completely covered by our field-of-view, according to the [OIII] maps shown
by \cite{soub89}. Its spectrum is noisier and with some clear defects at
specific wavelengths produced by the modeling technique. A detailed
inspection of the spectrum shown in Fig. 9 shows that the
[NII]$\lambda$6583 line is distorted by one of these defects, affecting the
\NA \ line ratio, which error is clearly larger than the formal error plotted
in Fig. 10.  Despite these caveats, it is clear that star formation
processes do not dominate the ionization of the different EELRs. 

The dust content in the EELRs is lower than the mean obscuration of the host. In particular,
for the E$_2$ and E$_3$ regions the dust absorption is almost half of the
average in the host galaxy. On the other hand, the electron density in those
clouds is $\sim$10 times lower than the average. However, this density is
$\sim$10 times higher in the E$_1$ region, i.e. $\sim$100 times larger than in
the other clouds.  This indicates most probably a different origin of those
clouds. In \cite{sanc04c} we discussed in detail the nature of the E$_1$
emission line region. It is most probably associated with the radio jet that
crosses this cloud, compressing it due to its lateral expansion, and splitting
it in two different kinematics regions \citep{axon89,sanc04c}.  The
compression is reflected in the increase of the electron density. The nature
of the ionization is not clear in this cloud, since a post-shock zone can also
give rise to the observed line ratios. The similarities between the line
ratios for the different EELRs, and between them and the host galaxy ones, may
indicate a similar origin for the ionization. That is, direct photo ionization
from the UV-field of the AGN. In that case the effect of the jet over the
E$_1$ would be reduced to a split of the cloud and a compression that give
rise to a density enhancement.

\subsection{Spectral energy distribution of the continuum dominated condensations}
\label{cont_sed}

Figure 9B shows the spectral energy distribution (SED) of the
continuum dominated structures A, B and C detected in 3C~120 within the
field-of-view of our IFS data. The SEDs were obtained by an average of the
integrated spectra of the different structures extracted from the residual
data cube over spectral ranges of 300\AA\ width. The low intensity of the
continuum dominated structures and the subsequent low signal-to-noise prevents
us of using directly the extracted spectra for this analysis. The SEDs were
cut at 7500\AA\ since at larger wavelengths the imperfect subtraction of the
sky-lines and the noise enhancement strongly affect the reliability of the
derived SED. Synthetic spectra for three different stellar populations of
different ages were included for comparison purposes. These spectra are
similar to those shown in the figure 8B (described in \S5.2),
but without including the effects of dust. The SEDs are almost flat in the
plotted ranges, being roughly consistent with a mix of a young stellar
population with an underlying old stellar population. This result agrees with
the results based on the broadband colors of the structures, discussed in
\S4.2.  Therefore, the continuum structures are experiencing a decrease of
the dust content, rather than an increase of the star formation. This seems to
be particularly valid for the A structure, which lies in a minimum of the dust
content, derived from the \AB\ line ratio (See Fig. 6E).
Unfortunately, we cannot check it for the B and C structures, because the
gaseous emission in those regions is less luminous and therefore the \AB\ 
ratio is uncertain.

So far only starlight was considered to explain the emission found in the
different condensations. However their SEDs could be also explained
considering other components, like scattered-light from the nucleus and/or
synchrotron emission associated with the radio-jet. \cite{hjor95} detected
polarization in the condensation $A$, which direction and magnitude were
consistent with those found by \cite{walk87}. This may indicate a connection
between that condensation and the radio jet. Indeed, it may indicate that
condensation $A$ consists, at least partially, of optical synchrotron
emission.  However, as it was shown in \S4.1 \citep[and noticed by][]{hjor95},
$A$ does not follow the radio-jet into the core, being more likely associated
with $B$ and $C$. Furthermore, the measured polarization may be due to
unsubtracted scattered-light from the nucleus \citep{hjor95}, that we detected
in the average spectrum of the host (Fig.  8 and \S5.2). Indeed,
the SEDs of the three different condensations do not differ significantly (Fig.
9B), and, in particular, they show almost the same $V-R$ colors
(Fig. 5), which indicates, most probably, a similar origin for
all of them.  Since most of the scattered-light has been removed
and the synchrotron radiation could also contaminate $A$, these condensations
are most probably dominated by starlight.

\section{3C~120 Velocity Field}

The central wavelengths of the Gaussians fitted to the individual spectra
(section \S4.3) give us the radial velocity associated to the ionized gas at
each position.  Interpolating the individual emission line centroids, we
obtain the velocity field of different emission lines. Uncertainties due to
wavelength calibration ($<$ 35 km s$^{-1}$) and those related to the fitting
process ($\sim$ 15 km s$^{-1}$) are small enough to be irrelevant for the
following discussion.

Figure 11 shows the velocity field of the ionized gas derived
from [OIII]$\lambda5007$ and H$\alpha$ lines. While H$\alpha$ traces the
kinematics of low ionisation gas that describe the general pattern of the
galaxy, [OIII] draws the signatures of high ionization gas, characterizing the
most perturbed regions. Previous kinematics studies of 3C~120 found a rather
chaotic velocity field \citep{bald80} and some slight evidences of a
co-existing rotating system \citep{bald80,mole88}. Despite of the different
spatial resolution and coverage of the IFS data in this paper, our results are
in agreement with those of \cite{bald80} and \cite{mole88} pointing towards
highly distorted kinematics in this object.

The unclear morphology and highly distorted gas kinematics of 3C~120 makes
a simple interpretation of the velocity structure difficult.  The ionized gas
velocity field presents a general regular pattern, with larger velocities at
the north-west and lower at the south-east that may be consistent with a disk
rotating around an axis along the north-east direction (figure
11A and 11B). Several kinematical perturbations
can be identify in the velocity maps as well as a regular rotation.

A velocity gradient is located at $\sim$5$\arcsec$ west of the nucleus,
aligned with the radio-jet (figure 3B), at the position of
E$_1$ (figure 11A). Using a better spectral
resolution, \cite{axon89} show that there are two different kinematics
components rather than a velocity gradient.  The north component is receding
while the south component is approaching. The kinematical perturbation is most
probably due to the lateral expansion of the radio-jet \citep{axon89}: as
already quoted, the interaction of the radio-jet and the intergalactic gas
produces an enhancement of the gas density (section \S4.3), and perturbs the
kinematics \citep{sanc04}.

The velocity field derived from [OIII] is much more distorted to the east
(farther than 4 arcsec from the nucleus) than that one derived from H$\alpha$.
Although the signal to noise in the IFS spectra is low and uncertainties are
larger in that region, differences in the kinematics may be explained by the
different origin of these lines. The comparison of the continuum dominated
structures and the velocity field derived from the H$\alpha$ shows that $A$
and $B$ coincide with a velocity distortion located at the west of the nucleus
(figure 11B).  This kinematical feature is similar to those
found in the velocity field of spiral galaxies because of the arms
\citep[e.g.][]{knap04}. This gives support to the idea of identifing the $A$
and $B$ regions as structures in an spiral arm in 3C~120 in spite of its
poor gas content and continuum emission dominated by starlight.  We will
discuss in detail the origin of this arm-like structure in section \S7.

The comparison of the ionized gas structures E$_2$, and E$_3$ (figure
7B) and the velocity field gives clues to the origin of the EELRs
(figure 11C). The morphology of E$_3$ follows remarkably well an
east-west gradient in the velocity map. E$_2$ is in a close to constant velocity
region. Both clouds are in regions which present strong
kinematical perturbations from the canonical rotation. These perturbations are
more prominent in the velocity structure of the high ionization gas than that
of the low ionization gas.

Fitting a single Gaussian to the emission lines in the residual data cube of
spectra (section \S5), we have derived the velocity behavior of the E$_1$,
E$_2$, and E$_3$ residual structures. Figure 11D shows the
velocity map derived from the [OIII] lines in the residual spectra. We will
refer to this map as the velocity map of residuals hereafter. The kinematics
of the structures trace remarkable well the perturbations in the velocity
field, although their kinematical features are smoother in the latter. The
smoothing is a clear consequence of the blending of the host galaxy and the
structure spectra. The velocity map of the residuals represents much better
the kinematics of the gas clouds in E$_1$, E$_2$, and E$_3$. We found a clear
gradient of velocities in the region corresponding to the three structures
pointing to inflows/outflows of gas at different inclination angles and
outside of the galactic disk.  While regions E$_1$ and E$_2$ present a
north-south velocity gradient, E$_3$ shows an east west gradient in good
agreement with the [OIII] velocity map (figure 11A).

Despite of the bulge dominated morphology found in the inner regions of
3C~120, the velocity field derived from H$\alpha$ indicates the presence of
a rotating disk along the north-west/south-east (figure 11B).
In that case, the receding velocities at the northwest suggest that southeast
is the face closer to the observer. The velocity gradient also traces the line
of zero velocities, but it is not clearly defined in the derived velocity field
 because of the distortions in the center introduced by the Sy1
nucleus. With a larger spatial coverage, \cite{bald80} determined a
PA=72$^{\circ}\pm15^{\circ}$ for the minor kinematics axis.  This angle agrees with
the semi-minor axis determined from the 2D modeling of the galaxy (section
\S4.1). A visual inspection of the velocity fields in figures
11A and 11B indicates that this PA may be a good
estimation of the minor kinematics axis, when considering the distortion at
the west. However, this distortion is most likely produced by the interaction of
the radio-jet with the E$_1$ cloud \citep{sanc04}, and most probably is not
related with the rotating disk. The isovelocity contours of the velocity field
derived from H$\alpha$ and the position of the kinematical center ($\sim0.8"$
north-west of the nucleus) suggest that the minor kinematical axis is most
likely along PA$\sim50^{\circ}$, which is not far from the photometric
determinations and within their uncertainties. Therefore, we will
consider hereafter a PA=$-40^{\circ}$ for the major kinematics axis of 3C~120. 

Figure 12 shows the rotation curve of 3C~120 derived from
the velocity field of H$\alpha$, assuming a PA of $-40^{\circ}$ for the major
kinematics axis. This curve is clearly distorted at the south-east due to the
kinematical perturbations associated with E$_2$. However, the north-west
portion of the curve is remarkable similar to those of galactic rotating disks
\citep{binn98}. The effects of the perturbations in the south-east portion of
the curve must be removed prior to model the 2D distribution of the rotating
component of the galaxy.  For doing so, it was assumed that the approaching
portion of the rotation curve (south-east) follows a symmetrical counterpart
of the receding portion (north-west). This {\it hypothetical} rotation curve
was included in Figure 12. This curve was then used to
derive a template of the velocity field by applying a simple rotational model
\citep{miha81}, for different inclinations varying from $\sim$5$^{\circ}$ to
$\sim$85$^{\circ}$ .  Subtracting those templates from the H$\alpha$ velocity
field and minimizing the differences at the northwest region (where the
rotation curve really corresponds to the 3C~120 kinematical behavior), we
estimated an inclination angle of $\sim40^{\circ}$ for the disk component of
3C~120. The velocity field template corresponding to that inclination angle is
shown in figure 11E. This estimation of the inclination is in
agreement with the determination from the external isophote \citep{mole88}.  As
quoted before, the internal isophotes are less elongated and distorted at the
west than the external ones. This suggests a different inclination angle,
smaller at the inner than at the outer regions. Indeed, the ellipticities
derived from the 2D modeling (section \S4.1) indicate that the inclination
may range from $\sim20^{\circ}$ to $\sim43^{\circ}$ in the center, in
agreement with the kinematical estimation.  The differences in inclination
from the center to the outer regions, although small, suggest a slightly warped
disk component in 3C~120.

Figure 11F shows the residual of the [OIII] velocity map once
subtracted the template shown in Fig. 11E (residual velocities
hereafter). This residual map shows an almost flat area of 0$\pm20$ \kms at
the north-west region, within the uncertainties of the velocity determination.
At the west, coincident with the path of the radio-jet and the location of
E$_1$, there is a north-south gradient of $\sim148$ km/s, in agreement with
the results by \cite{axon89}. At the location of E$_2$ there is a rather
chaotic residual velocity structure, with velocities ranging from $\sim$270 to
460 km/s. A slight east-west gradient of $\sim32$ km/s is found at the
location of E$_3$.
The residual velocities (Fig. 11F) and the velocity map of the
residuals (Fig. 11D) present a rather good qualitative agreement
in spite of the strong conceptual and practical differences of the two methods
used to derived them. This supports the idea of the presence of a rotational
component in 3C~120.

Due to the low spectral resolution of our data no attempt was done to
analyze the velocity dispersion maps.

\section{discussion}

In previous sections we presented several aspects of 3C~120 that describe a
puzzle environment. 3C 120 is morphologically speaking a bulge dominated
galaxy, which contains a rotating stellar disk and several continuum-
dominated structures and EELRs. 
We analyzed in detail those structures in order to determine their nature. The $E_1$ is caused by the interaction of the radio-jet with the intergalactic
medium, as already discussed in \cite{sanc04c}.

  The newly reported shells in the central region of 3C 120 is an
 addition to the complex picture of this object.
 Shells are common structures in elliptical  and SO galaxies, although a few 
 number of spirals also show the presence of this kind of features
 \citep{malin83,seitzer88}. Although models 
considering internal shock waves were proposed to account for the
 presence of shells in galaxies \citep[see e.g.][]{williams85}
 the most accepted idea is that they are generated by merging processes \citep[see e.g.][and reference therein]{hern92}. Numerical simulations indicate that the number and
 sharpness of shells in merger remnants depend on the mass ratio and the
 absence (or presence) of a central buldge  in the progenitors \citep{gonz04}.
The radial distribution
 of shells depends on the potential of the host galaxy and their morphology
 can appear aligned or randomly distributed around the galaxy \citep{prieur90}.
 3C 120 shows two well defined shells, labelled as S$_N$ and S$_S$, and
 some
 other shell-like faint features which are aligned along an axis of
 PA$\sim-25^{\circ}$. When shells are well aligned along a certain axis, 
this axis is always close to the major axis of the galaxy \citep{prieur90}, which is out case.
 Shells can be found in a wide range of radii from the nucleus, from only a 
 few to hundreds of Kpc, such us in the case of NGC 3923 \citep{prieur88}. The
 innest shell detected in 3C 120 is at around 1 arcsec north ($\sim 1 $ kpc)
 of the nucleus. Shells found close to the nucleus
 implies that a dissipative process has played an important role in the
 formation of the shells \citep{prieur90}. Therefore, the presence of the detected
 circumnuclear shells (section \S4.1) strongly suggests the idea of
 considering 3C 120 as a late stage merger.

 The arm-like clumpy structures A and B (section \S4.1) show kinematics that
 resembles those of a spiral arm as explained in section \S6. However, spiral
 arms are active star forming regions, and no star formation was detected in
 either of the continuum-dominated structures. In that case, these structures
 could be also the remnant of a merging process. Indeed, the regions B, A and
 the S$_s$ seem to be connected in our residual and color maps (Fig 3C and 4B),
 and S$_s$ is associated with the E$_2$ EELR. A kinematics feature (Fig. 11B)
 is coincident with these structures. Numerical simulations indicate that these
 kind of structures are found as remnants of merging processes 
 \citep[e.g.][]{howa93,heyl96,miho96} being difficult to be explained by any other mechanism. Assuming that scenario, the presence of
 an over density of gas in the south shell may indicate an inflow, where the
 warmer gas has dropped faster into the inner regions. This picture is in good
 agreement with the kinematics features at the south shell (E$_2$ clouds
 kinematics) and its residuals. The regular velocity in this region suggests
 that this inflow may occur in a plane outside the galactic disk and close to
 the line of sight. Although the kinematics of the merger remnants
 strongly depends on the viewpoint, the velocity and velocity dispersion
 measured for E$_2$ is in good agreement with merger simulations
 \citep{heyl96}. Theoretical models indicate
 that inflows channeling gas
 from the outer to the nuclear regions may appear in the late stages of a
 merging process \citep[e.g.][]{miho96}.  These inflows have been already
 observed in different merging galaxies \citep[e.g.][]{arr02,arr03}.
  The discussion above suggests that the structures A, B, and the E$_2$ EELR
 can be identified with a tidal tail driving material to the central region most probably due to a past merger event in 3C 120. Most of the analyzed continuum dominated 
structures shows a rather young stellar population, indicating a recent 
(but not ongoing) star formation rate. This result agrees with the picture 
of a post-merging event. Similar results have been found in the study of 
the optical colors of host galaxies AGNs \citep{jahnke04,sanc04b} .

 The poor spatial coverage of the third identified EELR, E$_3$, prevents us of
drawing a clear picture of its origin. In fact, this region corresponds (see
section \S5.1) to a tail extension of a larger structure already reported in
this object \citep{soub89}. Although we cannot be conclusive with our current
data, the kinematical results most probably indicate that E$_3$ is associated
with an inflow/outflow of gas to/from the circumnuclear region.

It has long been suggested that strong interactions and galaxy mergings may
(re-)ignite nuclear activity \citep[e.g.][]{sand88}. Galaxy interactions can
produce the loss of momentum required to allow the infall of gas towards the
nuclear regions, gas that would feed the AGN. Many authors found that AGN
hosts show distorted morphologies, reminiscent of past merging events
\citep[e.g.][]{mcle94a,mcle94b,bahc97,sanc03b}. They are found in environments
with high probability of experiencing interactions \cite[e.g.][]{sanc03a}. And
their hosts, mostly early-type galaxies, present anomalous blue colors
\cite[e.g.][]{sanc04c}. All together it may indicate that, if not all, at
least a family of AGNs is generated by the merging/interaction between
galaxies \citep{cana01}. A merging event alone id not enough to
generate an AGN. The presence of a massive black-hole in the progenitor galaxy
is a basic requerement. Massive black-holes are only found in bulge-dominated
massive galaxies, due to the black-hole/bulge mass relation
\citep[e.g.][]{mago98,korm01}. Indeed, recent results (S\'anchez et al., in
prep.) show that the fraction of AGNs increases in galaxy mergers between two
large galaxies or a large with an small galaxy. 

Our current results agree with this sceneario. 3C~120 is a bulge dominated
galaxy which has, most probably, experienced a merging event with a less
 massive galaxy. That galaxy was completely disrupted in the merging process,
 falling in parts which produce, most probably, many of the observed structures
\citep{mole88}. This may explain the different stellar populations of the
structures and the average stellar population in the object (\S4.2) as
 it has been previously reported for other galaxies \citep{prieur90}. A
subtantial fraction of its gas has been channelled towards the inner regions,
following the detected arm-like structure, and concentrating in the E$_2$
region. 

\section{Summary}

We obtained integral field optical spectroscopy of the Seyfert 1 radio galaxy
3C~120.  The homogeneous data, excellent spatial and spectral coverage, and
good spatial and spectral resolution make this atlas a useful tool for
studying 3C~120 in the optical. These IFS optical data were combined with high
resolution HST imaging. The analysis of these data suggests that a Seyfert 1
 nucleus at the center, an early type galaxy, and several structures formed
 as a consequence of a merging process in the past history constitute the
 radio galaxy 3 120. At least one of these
 structures is identified with an inflow, which is feeding the central engine.
 A radio-jet is escaping from the center, perturbing the gas on its path.

The main results from the analysis of this dataset were the following:

\begin{itemize}
  
\item[1.] Several continuum-dominated structures were detected in 3C 120,
  which do not follow the mean distribution (bulge dominated) of the stellar
  component.  Some of these structures were shells in the central Kpc of the
  object, which may indicate a past merging process. The colors of the
  different components were not compatible with a single stellar population in
  this galaxy.
  
\item[2.] Three emission line structures were identified (E$_1$, E$_2$ and
  E$_3$) which were not associated with the general behavior of the galaxy.
  These gaseous structures presented a high level of ionization. The origin of
  structure E$_1$ was the interaction between the intergalactic medium and the
  radio-jet emerging from the nucleus.
  
\item[2.] The spectra of the nucleus and the host galaxy were decoupled,
  obtaining a data cube of residuals. From that residual data cube, we
  extracted and analyzed the spectra of the different structures in 3C 120.
  
\item[3.] The velocity field indicated a rotational component plus several
  kinematical perturbations associated with the identified emission
  structures.

\item[4.] The continuum-dominated structures A, B, the S$_s$ shell and the
  EELR E$_2$ seem to be physically associated, belonging to the inner-most
  part of a tidal-tail, remanent of a past merging event. This tail has
  channeled gas into the inner regions, in an inflow, that has generated
  E$_2$.

\end{itemize}

The 4.2-m William Herschel Telescope is operated by the Isaac Newton Group at
the Observatorio de Roque de los Muchachos of the Instituto de Astrof\'\i sica
de Canarias. We thank all the staff at the Observatory for their kind support.
This project is part of the Euro3D RTN on IFS, funded by the EC under contract
No.  HPRN-CT-2002-00305. This project has used images obtained from the HST
archive, using the ESO archiving facilities. We would like to thank Dr.Walker
that has kindly provided us with the radio maps of 3C~120.


\clearpage

\appendix

\section{Atlas of spectra}

In this appendix we present the spatial distribution of the brightest emission
line profiles as spectra diagrams. These diagrams represent line profiles in a short wavelength range of the individual spectra at each point (fiber). The spectra at each location are autoscaled to show the profile shape (lines nearer to the optical nucleus are brighter than those farther out). Figure 13 and 14 show the spectra diagrams corresponding to H$\beta$+[OIII]$\lambda\lambda4959,5007$, and H$\alpha$+[NII]$\lambda\lambda6548,6584$+[SII]$\lambda\lambda6716,6730$, respectively. A considerable blending of the emission lines with the wings of the broad component of the Balmer lines can be clearly appreciated in these diagrams. 

\section{Decoupling the spectra of different components using IFS}

A new technique to decouple the spectra of different components in an object
has been developed, based on well known techniques applied in the decoupling
of different components in images. Two different methods have been developed.
A first one is based on a 2D modeling of the objects using object templates,
while the second one is based on an isophotal analysis of 2D images. We
explain here how both methods have been extended to the analysis of IFS
datacubes, and the advantages of each one.

\subsection{3D modeling}
\label{ap_deb}

IFS combines the characteristics of imaging and spectroscopy. In general
terms, IFS data consist of a series of spectra obtained at a discrete number of
positions in the sky. In the case of INTEGRAL data in the SB2 configuration,
they consist of the 209 spectra obtained in the same number of positions in
the sky. Newer instruments, like PMAS \citep{roth00} or VIMOS \citep{lefe03},
sample the sky in a continuous way, by coupling the fibres to lens-arrays. In these latter cases
a IFS datacube can be understood as a stack of narrow-band images.
More generally, at each wavelength we have a 2D discrete distribution of
fluxes, that can be transformed to a regular grid distribution by an
interpolation routine.

It is a quite common situation that an observed object is the combination of
different components: e.g., an QSO lens, a crowded-field of stars and an
AGN+host system. Different techniques have been developed to decouple the
different components of an image, and to derive their photometry. A wide
extended technique is to fit the images with a 2D model, including a template
for each different component. This technique is implemented in GALFIT
\citep{peng02}, a program for modeling multi-components in images. This
program has been used in the decoupling of nuclear and host components
\citep{sanc04b} and the decoupling of different components in a QSO lens
\citep{wiso04}. GALFIT fits the images with a multi-component model, producing
a template image and a residual image. This residual image can be used to
detect substructures in 2D images.

A natural extension of the 2D modeling technique to IFS is to split the
datacube in a set of narrow-band images of the width of the spectral pixel
(the so-called monochromatic images), and treat them as individual images.
This technique has been used successfully for the deblending of QSO lenses
with IFS \citep{wiso03}. In the particular case of our INTEGRAL data, we first
transformed the discrete set of spectra in a regular grid datacube using an
interpolation routine implemented in E3D \citep{sanc04}. The final datacube
has a 0$\farcs$3$\times$0$\farcs$3 pixel size, i.e., 1/3th of the original
diameter of the fibers.

The 2D image modeling of the nucleus and the host for each monochromatic
image was performed using GALFIT. The 2D model comprises a narrow Gaussian
function (to model the nucleus) and a de Vaucouleurs law (to model the
galaxy), both convolved with a PSF. The PSF at each wavelength was obtained
from a calibration star datacube, observed just before the object. In a first
attempt all the parameters of the model were fitted: the intensities of the
nucleus and the host, the centroid of the object, the effective radius, the
position angle and the ellipticity of the host galaxy. As already noticed by
\cite{wiso03}, this first modeling does not produce acceptable results. The
large number of parameters, the degeneracy between some of them (e.g., the
total intensity and the scale-length), and the limited sampling of our data
impose limitations to the quality of the recovered spectra.

However, \cite{wiso03} demonstrated that it is possible to increase the
quality of the modeling imposing certain conditions in the structural
parameters. The centroid of the object varies through the wavelengths due to
differential atmospheric refraction (DAR) \citep{fili82}. 
Therefore, whenever there is a clear peak in the object (like in our case) the
centroid should vary smoothly with the wavelength. Figure 15 shows
the centroid coordinates as a function of the wavelength resulting from the 3D
modeling. Despite of the fluctuations, there is a clear trend.  The
solid-lines show the result of a fitting a polynomial function of order
5. In a second run of the 3D modeling we imposed the centroid to be fixed to
the result from this polynomial fitting. Similar conditions can be imposed
for the ellipticity, effective radius and position angle of the host galaxy.
By forcing the structural parameters to vary smoothly with the wavelength and
fixing them in the 3D modeling process the number of parameters is reduced to
the intensity of the different components. Thus the accuracy of the modeling
is increased \citep{wiso03,wiso04}.

The number of parameters can be reduced too if we derive the structural
parameters from an external source of information.  As we explained above we
have good quality HST images of the object in different bands, that have been
already analyzed using GALFIT (\S4.1). We fixed the structural parameters
(effective radius, ellipticity and position angle) in our second 3D modeling
of the values derived by the 2D modeling of the HST images (Table
\ref{tab_galfit}). The centroids were fixed to the values derived using the
polynomial function shown in Fig. 15.  Figure 16
illustrates the process. The top-left panel shows the decoupled spectra of the
nucleus and the host derived from the 3D modeling when all the parameters are
fitted freely.  The spectra derived from the first run are noisy and can only
be considered as a very rough approximation. The top-right panel shows the
spectra once fixed the centroids to the result of the polynomial function
fitting shown in Fig.  15. There is a clear increase of the quality
of the derived spectra.  The bottom-left panel shows the spectra when, in
addition to the centroids, the structural parameters of the host galaxy
(ellipticity, position angle and effective radius) are fix to the result from
a polynomial function fitting similar to the one used for the centroids. This
is the best possible result if we only had the information from the IFS.  The
bottom-right panel shows the final result, obtained when the structural
parameters are fixed to the values derived from the 2D modeling of the HST
images. The derived spectra are remarkable similar to those of the previous
case.  However, in the previous case it is possible to identify some {\it
  fake} absorption features in the spectrum of the nucleus coincident with the
narrow emission lines (e.g., [OIII]$\lambda$5007). We cross-checked the
accuracy of the final decoupled spectra by comparing the fluxes derived with
that obtained by the 2D modeling of the HST images. Both fluxes agree
within $\sim$20\%.

As we explained above GALFIT provides us with a residual image at each
wavelength once subtracted the 2D model from each monochromatic image.  We
stored and stacked those residual images to create a residual datacube. The
residual datacube is used (1) to cross-check the accuracy of the fitting,
i.e., no evident fake residuals are found, and (2) to study the substructures
in the object. Figure 17, left panel, shows a narrow-band image
centred in the continuum adjacent to the [OIII] emission line (5204-5246\AA)
from the residual datacube (contour-plots), together with the residual image
obtained by a 2D modeling of the HST F555W-band image (grayscale). A simple
comparison of both images shows that the residuals are remarkable similar
outside $\sim$2$\arcsec$ from the nucleus. However, there is a ring structure
at $\sim$1$\farcs$5, $\sim$100 fainter than the nucleus, expecially bright in
the S$_S$ of the nucleus. This structure coincides somehow with the south-east
shell.  However it is worrisome that it is fainter in the north-west, where the
S$_N$ shell is brighter. Taking into account the original size of the fibers
($\sim$0$\farcs$9) and our limited determination of the PSF, we think that
this ring substructure is an artifact of the 2D modeling. Although it limits
the use of the residual datacube to study the properties of the substructures,
it does not significantly affect the accuracy of the decoupled spectra.

\subsection{3D surface brightness analysis}

We developed a second method to create a 3D template of the object that would
not depend on the PSF determination and without assuming any certain law to
describe the intensity profile. This method is an extension of the isophotal
surface brightness analysis technique \citep{jedr87}. It looks for the ellipse
parameters that better represent the isophotal shape by a Fourier analysis of
the surface brightness at a certain radius as a function of the eccentric
anomaly. The output is a surface brightness profile of the object (i.e., the
radial distribution of the intensity), and the radial distributions of the
centroid, position angle and ellipticity (understanding radius as semi-major
axis of the ellipse). A template of the object is created using these output
parameters. Then, a residual of the image is obtained by subtracting this
template from the original image.  This technique has been extensively used
for the detection of substructures in galaxies for decades
\cite[e.g.][]{pere86,hjor95}.

We extended this method to analyze substructures in IFS data. For each
monochromatic image we performed an isophotal surface brightness analysis
using our own coded routines. The average centroids, ellipticities, position
angles, and the integrated fluxes obtained from this analysis are stored for
each wavelength. The residual images obtained are combined in a residual
datacube. Like in the previous method the quality of the model template can be
increased assuming that the structural parameters change smoothly with the
wavelength. The structural parameters for each wavelength were fixed to the
values derived by a polynomial function fitting. In the case of an isophotal
analysis, once fixed the structural parameters, the derived model for each
monochromatic image is just estimated obtaining the average flux at a certain
radius along the eccentric anomaly. I.e., it is not a fit anymore, but just a
direct measurement. 

Figure 18 illustrates the results of this process. It shows, in
both panels, the integrated spectrum of 3C~120 over the field-of-view of
INTEGRAL (black line), together with the recovered spectrum from the 3D
template (orange line), and the integrated spectrum of the residuals. The left
panel shows the result from the first iteration where all the structural
parameters are derived by the described method. In general terms the spectrum
of the model matches well the original spectrum, despite some
wavelengths where the program is unable to derive the correct parameters.  The
right panel shows the result from the second iteration with all the structural
parameters fixed, as described above. In this case the agreement between the
recovered and the original spectrum is remarkable good, with no flux deviation
higher than a few percent at any wavelength.

As expected this method produces a better quality residual datacube than
the 3D modeling. The basic reason is that we reduced the number of assumptions
about the shape of the object and we transformed a fit to a direct measurement.
Figure 17, right panel, shows a similar narrow-band image of the
continuum adjacent to [OIII] from the residual datacube than the one presented
in the left panel, but obtained with the 3D surface brightness analysis.  The
grayscale shows once more the residual from the 2D modeling of the F555W-band
image. The residuals from the narrow-band image and the HST data match better
than that of the 2D modeling. No trace of the ring structure is present,
which reinforce our idea that it is an spurious result due to a limited
sampling and inaccuracies in the PSF determination.  

The two developed techniques to decoupling spectra of different components
from IFS data give successful results. The first method could be useful to
deblend spectra, such in the case of spectra from different components. The
second procedure is more suitable to study the residual structures once the
spectrum of the main component is subtracted.

\clearpage


\begin{table} 
\begin{center}
\caption{Summary of properties of the HST images}
\label{tab_hst}
\begin{tabular}{lcccr}\hline
\tableline\tableline
Band & Exp.Time & 3$\sigma$ lim. mag & mag & mag\\
 & (s) & (mag/arcsec$^2$) & & NED\\
\tableline
F555W (V) & 2040 &  25.3 & 14.94& 14.7\\
F547M (V')& 2300 &  25.1 & 14.35& ----\\
F675W (R) & 2250 &  25.3 & 13.74& 13.9\\
F814W (I) & 2236 &  24.9 &  (1) & 12.9\\
\tableline
\end{tabular}

(1) Peak saturated.

\end{center}
\end{table}
\begin{table} 
\begin{center}
\caption{\label{tab_galfit}Summary of the result from the 2D modeling}
\begin{tabular}{ccccccc}\hline
\tableline\tableline
Band & $n$ & m$_{\mathrm{nuc}}$ & m$_{\mathrm{host}}$ & r$_{\mathrm{e}}$ & a/b
& PA($^o$)\\
\tableline
F555W & 4.7& 16.1 & 15.4 &  1.9 kpc&0.87&$-$16.8\\
F547M & 4.5& 16.9 & 14.5 &  2.9 kpc&0.86&$-$18.6\\
F675W & 6.7& 16.1 & 13.9 &  2.8 kpc&0.73&$-$19.1\\
F814W & ---& ---- & 13.2 &  -------&0.94&$-$18.5\\
\tableline
\end{tabular}
\end{center}
\end{table}

\begin{table} 
\begin{center}
\caption{Properties of the emission lines detected in the nucleus and host galaxy of 3C~120}
\label{elines1}
\footnotesize
\begin{tabular}{llrrrr}\hline
\tableline\tableline
Name&Line & $\lambda$$^1$ & $\sigma_\lambda$ & Flux$^2$ \\
\tableline
Nucleus 
 &H$\delta$$^3$        & 4250.33 & 24.12       &   61.25 & 260.78     \\ 
 &H$\delta$BL1     & 4249.15 & 24.12       &  319.60 & 583.19     \\ 
 &H$\delta$BL2     & 4254.50 & 24.12       & 1754.62 & 516.10     \\ 
 &H$\gamma$        & 4498.42 & 24.12       &   35.08 & 268.68     \\ 
 &H$\gamma$BL1     & 4497.24 & 24.12       &  885.98 & 539.97     \\ 
 &H$\gamma$BL2     & 4502.58 & 24.12       &  790.73 & 422.00      \\ 
 &[OIII]$\lambda$4363    & 4521.98 & 24.12       &   60.99 & 119.60     \\ 
 &HeII$\lambda$4686BL1       & 4846.66 & 24.12       &  192.56 & 100.54     \\ 
 &HeII$\lambda$4686BL2       & 4876.41 & 24.12       &  270.39 & 141.18       \\ 
 &H$\beta$         & 5024.82 &  0.19       &   75.54 & 479.99     \\ 
 &H$\beta$BL1     & 5028.23 &  1.76       & 1473.31 & 754.88      \\ 
 &H$\beta$BL2     & 5058.85 & 53.77       & 1676.73 & 788.82      \\ 
 &[OIII]$\lambda$4959    & 5125.77 &  0.19       &  163.17 &  35.06     \\ 
 &[OIII]$\lambda$5007    & 5175.36 &  0.19       &  489.58 & 105.20     \\ 
 &HeI$\lambda$5876BL1   & 6075.86 & 11.62       & 332.05 & 112.13       \\ 
 &HeI$\lambda$5876BL2   & 6105.86 & 11.62       & 465.42 & 157.05       \\ 
 &[OI]$\lambda$6300.3    & 6518.52 & 40.42       &   20.34 &  33.63     \\ 
 &[OI]$\lambda$6363.8    & 6584.19 & 40.42       &   30.34 &  34.81     \\ 
 &[NII]$\lambda$6548     & 6784.48 &  2.08       &    9.46 &  57.34     \\ 
 &H$\alpha$        & 6799.71 &  2.08       &  232.62 & 513.60      \\ 
 &H$\alpha$BL1     & 6782.94 &  0.58       & 3570.51 & 574.07      \\ 
 &H$\alpha$BL2     & 6797.59 &  1.15       & 5985.15 & 446.94      \\ 
 &[NII]$\lambda$6583     & 6821.21 &  2.08       &   28.40 & 172.13     \\ 
\tableline
HOST 
 &[OII]$\lambda$3727     & 3867.81 &  10.22     &  477.51 &  60.96       \\ 
 &[NeIII]$\lambda$3869   & 4010.57 &  16.32     &  332.54 &  48.24       \\ 
 &[NeIII]$\lambda$3967   & 4117.10 &  16.32     &  179.77 &  34.32       \\ 
 &[SII]$\lambda$4072     & 4223.39 & 122.12     &  110.88 &  37.32       \\ 
 &H$\gamma$        & 4494.84 & 389.92     &   54.84 &  28.56         \\ 
 &[OIII]$\lambda$4364    & 4518.40 & 389.92     &   56.88 &  28.32        \\ 
 &HeII$\lambda$4686      & 4851.17 & 143.31     &  101.76 &  33.48         \\ 
 &H$\beta$         & 5030.04 &   0.82     &  218.05 &  39.72            \\ 
 &[OIII]$\lambda$4959    & 5131.00 &   0.82     & 1164.08 & 161.37     \\ 
 &[OIII]$\lambda$5007    & 5180.58 &   0.82     & 3495.62 & 484.59     \\ 
 &HeI$\lambda$5876       & 6072.99 &  23.27     &   45.60 &  47.40     \\ 
 &[OI]$\lambda$6300      & 6514.63 & 194.46     &  122.40 &  60.48      \\ 
 &[OI]$\lambda$6364      & 6580.30 & 194.46     &   58.08 &  55.20      \\ 
 &[NII]$\lambda$6548     & 6771.69 &   2.31     &  259.45 &  46.92     \\ 
 &H$\alpha$        & 6786.91 &   2.31     & 2097.63 & 255.97           \\ 
 &[NII]$\lambda$6583     & 6808.41 &   2.31     &  778.37 & 140.77    \\ 
 &[SII]$\lambda$6716     & 6946.13 &  32.68     &  368.18 &  96.00     \\ 
 &[SII]$\lambda$6731     & 6961.00 &  32.68     &  248.29 &  89.52    \\ 
 &ArIII         & 7377.62 & 291.60     &  108.48 &  69.00        \\ 
\tableline
\end{tabular}
(1) wavelength in \AA, (2) flux in \fluxunitsAA units, (3) the narrow-emission
    lines had a velocity dispersion of $\sim$700 \kms, while the
    broad-emission lines have to component, BL1 and BL2, with a velocity
    dispersion of $\sim$2188 \kms and $\sim$8706 \kms, respectively.

\end{center}
\end{table}

\begin{table}  
\begin{center}
\caption{Properties of the emission lines detected in the EELRs of 3C~120}
\label{elines2}
\footnotesize
\begin{tabular}{llrrrr}\hline
\tableline\tableline
Name&Line & $\lambda$$^1$ & $\sigma_\lambda$ & Flux$^2$ \\
\tableline
E$_1$ 
 &[OII]$\lambda$3727     & 3868.81 &  8.63     &  1.45 & 0.38        \\ 
 &[NeIII]       & 4015.92 &  8.63     &  0.13 & 0.02          \\ 
 &[NeIII]       & 4117.45 &  8.63     &  0.68 & 0.09         \\ 
 &H$\beta$         & 5032.50 &  3.74     &  2.11 & 0.17          \\ 
 &[OIII]$\lambda$4959    & 5133.57 &  3.74     &  9.52 & 1.21    \\ 
 &[OIII]$\lambda$5007    & 5183.20 &  3.74     & 28.55 & 3.64    \\ 
 &[NII]$\lambda$6548     & 6775.53 &  8.62     &  1.53 & 0.13     \\ 
 &H$\alpha$        & 6790.75 &  8.62     & 16.32 & 1.39         \\ 
 &[NII]$\lambda$6583     & 6812.25 &  8.62     &  4.62 & 0.40   \\ 
 &[SII]$\lambda$6716     & 6948.04 & 58.25     &  2.63 & 0.24   \\ 
 &[SII]$\lambda$6731     & 6962.91 & 58.25     &  2.06 & 0.21   \\ 
\tableline
E$_2$
 &[OII]$\lambda$3727     & 3864.80 &  10.25     &  7.08 & 0.55   \\ 
 &[NeIII]       & 4012.37 &  13.09     &  4.29 & 0.35    \\ 
 &[NeIII]       & 4111.74 &  26.17     &  2.53 & 0.20    \\ 
 &[SII]         & 4217.03 &  82.77     &  0.56 & 0.06    \\ 
 &H$\delta$     & 4247.03 &  10.82     &  1.02 & 0.11    \\ 
 &H$\gamma$     & 4493.72 &  10.82     &  2.48 & 0.32    \\ 
 &HeII$\lambda$4686      & 4850.62 &  40.48     &  2.20 & 0.17  \\ 
 &H$\beta$      & 5028.49 &   0.81     &  4.95 & 0.39    \\ 
 &[OIII]$\lambda$4959    & 5129.44 &   0.81     & 25.27 & 3.06   \\ 
 &[OIII]$\lambda$5007    & 5179.03 &   0.81     & 75.82 & 9.18   \\ 
 &HeI$\lambda$5876       & 6077.20 &  50.43     &  0.80 & 0.12   \\ 
 &[OI]$\lambda$6300.3    & 6513.68 &  40.26     &  1.63 & 0.20   \\ 
 &[OI]$\lambda$6363.8    & 6572.96 & 294.61     &  0.24 & 0.04   \\ 
 &[NII]$\lambda$6548.1   & 6770.28 &   3.38     &  2.77 & 0.23   \\ 
 &H$\alpha$        & 6785.51 &   3.38     & 25.40 & 2.04         \\ 
 &[NII]$\lambda$6583.6   & 6807.01 &   3.38     &  8.33 & 0.68   \\ 
 &[SII]$\lambda$6716     & 6946.47 &  16.71     &  5.89 & 0.50   \\ 
 &[SII]$\lambda$6731     & 6961.35 &  16.71     &  2.96 & 0.29   \\ 
 &ArIII         & 7378.28 &  51.65     &  1.48 & 0.12            \\ 
\tableline
E$_3$
 &[OII]$\lambda$3227     & 3866.04 & 14.72     &  4.54 & 0.87    \\ 
 &[NeIII]$\lambda$3927   & 4013.15 & 14.72     &  2.39 & 0.28    \\ 
 &H$\beta$      & 5032.51 & 39.54     &  2.10 & 0.21              \\ 
 &[OIII]$\lambda$4959    & 5131.26 &  2.71     & 11.06 & 1.53   \\ 
 &[OIII]$\lambda$5007    & 5180.84 &  2.71     & 33.21 & 4.58  \\ 
 &[NII]$\lambda$6548     & 6771.15 &  6.12     &  0.41 & 0.01  \\ 
 &H$\alpha$     & 6786.37 &  6.12     & 12.74 & 1.71  \\ 
 &[NII]$\lambda$6584     & 6807.87 &  6.12     &  1.23 & 0.04  \\ 
 &[SII]$\lambda$6716     & 6946.55 & 22.92     & 13.97 & 1.56  \\ 
 &[SII]$\lambda$6731     & 6961.42 & 22.92     &  6.70 & 0.75  \\ 
\tableline
\end{tabular}

(1) wavelength in \AA, (2) flux in \fluxunitsAA units.

\end{center}
\end{table}

\begin{table*}  
\begin{center}
\caption{Properties of the gaseous emission of the different components in 3C~120}
\label{param}
\small
\begin{tabular}{lccccccc}\hline
\tableline\tableline
Name&\OB&\NA&\Hab&$A_V$    &\Srat&$n_e$    &T$_{eff}$\\
    & $^1$  & $^1$  &    & mags  & &cm$^{-3}$& (K)\\
\tableline
E$_1$  & 1.13 & -0.54 & 7.7 & 3.2 & 1.3 & $\sim$160 & --- \\
E$_2$  & 1.18 & -0.48 & 5.1 & 1.9 & 2.0 & $\sim$1 & --- \\
E$_3$  & 1.20 & -1.02 & 6.1 & 2.4 & 2.0 & $\sim$1 & --- \\
Nucleus$^2$& 0.81 & -0.91 & 3.1 & -- & -- & -- & 2.3$\times$10$^5$\\
Host   & 1.21 & -0.43 & 9.8 & 4.0 & 1.5 & $\sim$10 &  1.4$\times$10$^4$\\
\tableline
\end{tabular}

(1) In logarithms. (2) Narrow emission lines

\end{center}
\end{table*}

\clearpage


\plotone{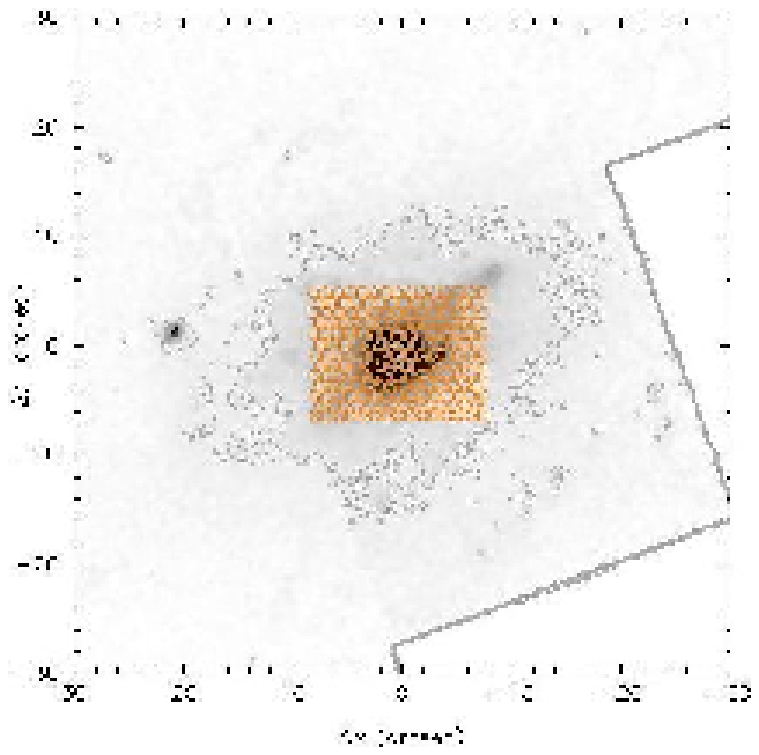}
{\bf Figure 1:} F555W-band image of 3C~120 taken with the WFPC at the HST.
The spatial distribution of the science fibers at the focal plane of the 4.2m WHT has been overlaid. North and East are up and left, respectively as usual. The orientation is the same in all figures in this paper.

\plotone{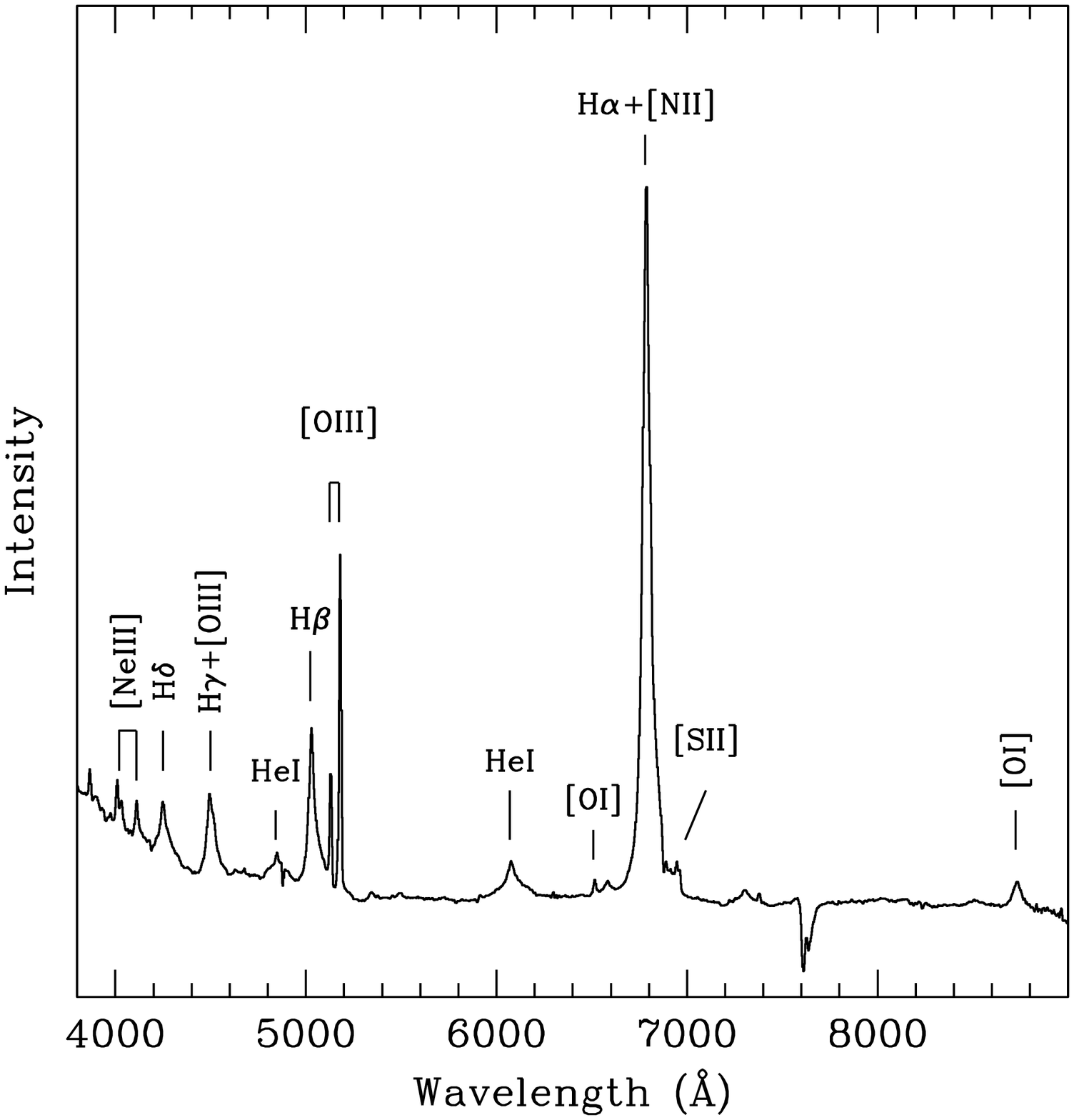}
{\bf Figure 2:} Nuclear spectrum in the full spectral range. This
spectrum corresponds to a hexagonal aperture of 1$\farcs$6 in radius centred
on the optical nucleus.

\plotone{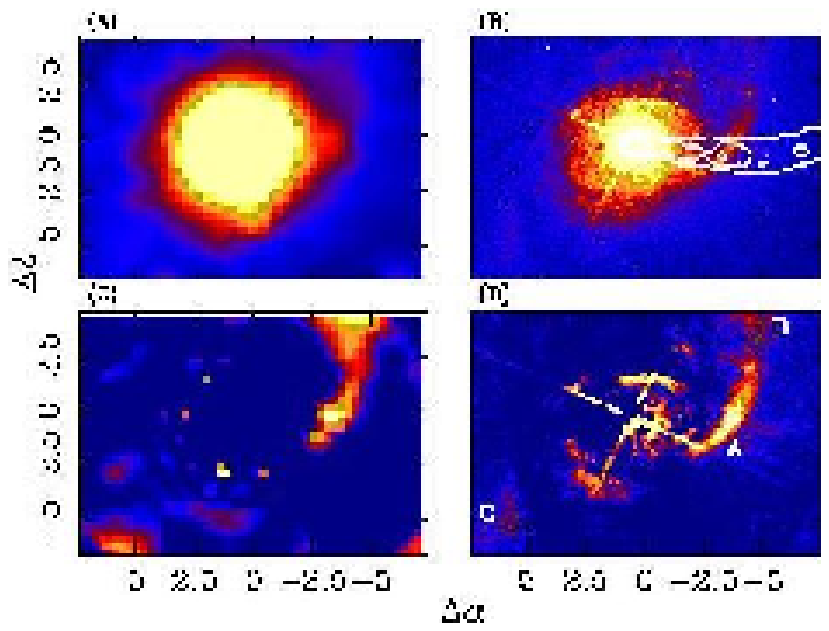}
{\bf Figure 3:} (A) Two-dimensional distribution of the continuum emission
of 3C~120 derived from the IFS data by integrating the signal in the spectral
range 5100-6100 \AA , similar to that range covered by the $V$-band filter.
(B) F555W-band image of the same central region of 3C~120 obtained with the
HST/PC (0$\farcs$046/pixel). Different continuum dominated structures already
detected from ground base observations are labeled using the nomenclature
from \cite{soub89}.  The contour-plot shows the radio map at 4885 MHz
\cite{walk97}. (C) $V$-band residual image, once subtracted the galaxy
template derived by a 2D modeling from the $V$-band image shown in panel (A).
(D) F555W-band residual image derived by subtracting a galaxy template
obtained by a 2D modeling from the F555W-band image shown in panel (B).

\plotone{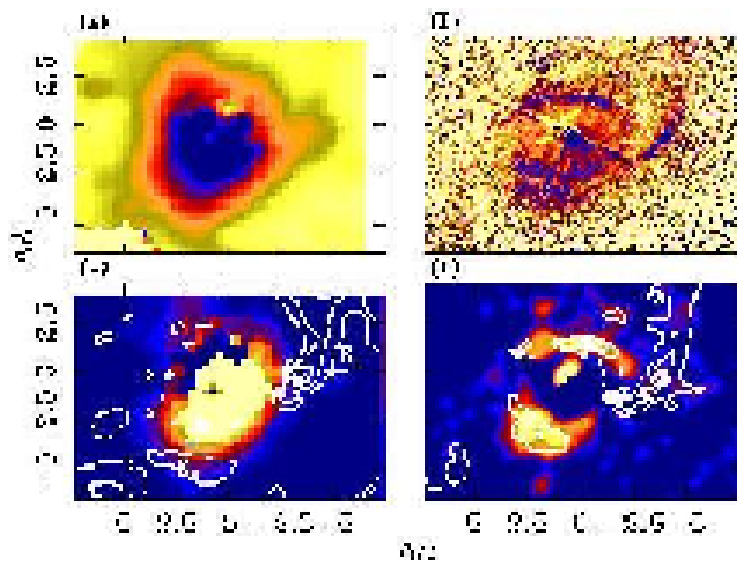}
{\bf Figure 4:} (A) The $V-I$ color image derived from the $V$ and $I$ maps
recovered from the IFS data. (B) The $V-I$ color image obtained from the F555W
and F675W-band HST images. The image has been smoothed with a 3$\times$3 pixel
median kernel. (C) The residual V map (close similar to an ionized gas map)
derived subtracting a continuum recovered from the IFS data (5600-5700 \AA )
avoiding the contribution of any bright emission line and scaled. (D) The
residual image obtained by the subtraction of the F547M-band image from the
F555W-band image, once scaled by the differences of the photometric
zero-points. The contour-plot corresponds to the F555W-band residual image
shown in figure 3D. Labels indicate the previously detected
EELRs \citep{soub89,sanc04c}. 

\epsscale{0.60}
\plotone{f5.ps}

{\bf Figure 5:}  $V-R$ colors as a function of the $R-I$ colors of the host
galaxy and the different structures found in 3C~120. The discontinuous line
shows the location of a single stellar population model at different ages
\cite{bruz03}. The values indicate the logarithm of the age in Gyrs. The
dotted line shows the average colors of different galaxy types (E, Sab, Sbc,
Scd and Irr), obtained from \cite{fuku95}. The arrow shows the effect of dust
in the colors for an absorption of A$_\mathrm{V}\sim$1.5 mags. Errors of 
the individual colors are 0.1 and 0.26 mag in V-R and R-I, respectively.

\epsscale{1}
\plotone{f6.ps}
{\bf Figure 6:}   Integrated intensity maps of (A) [OIII]$\lambda5007$
and (B) H$\alpha$ emission lines derived by a Gaussian fitting. In the
circumnuclear region, we have included a second broad Gaussian to H$\alpha$.
Dividing line intensity maps, we have derive: (C) the ionization map
[OIII]/H$beta$; (D) [NII]/H$\alpha$ 2D-distribution; (E) the extinction
H$\alpha$/H$\beta$ map; and (F) the electronic density distribution
[SII]$\lambda6716$/[SII]$\lambda6730$.

\epsscale{0.95}
\plotone{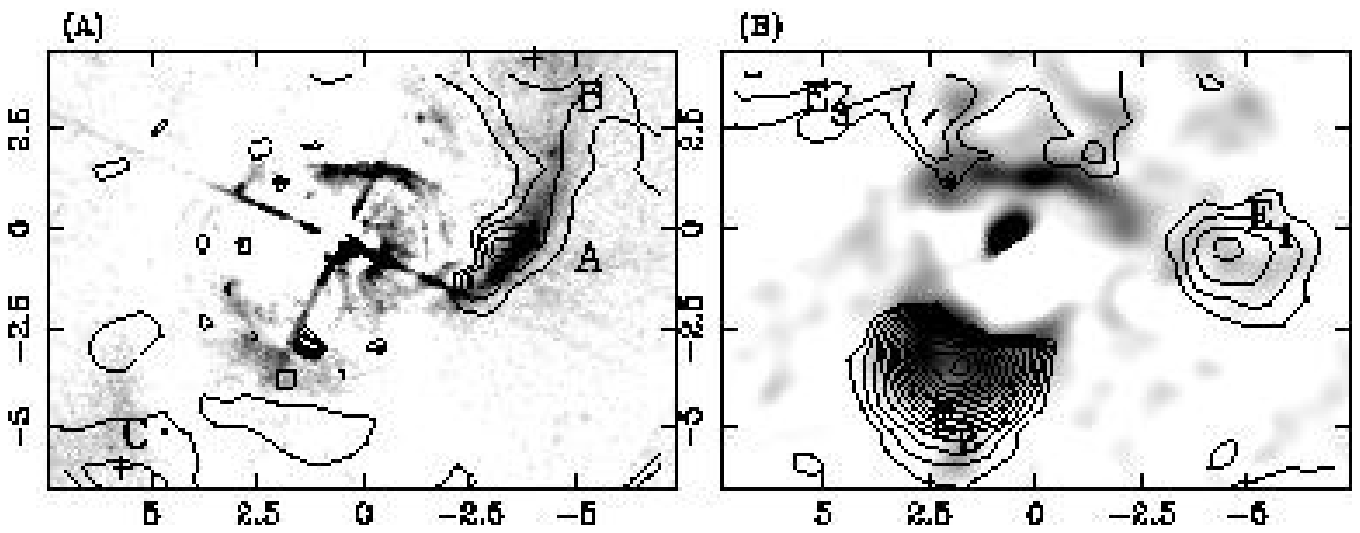}
{\bf Figure 7:} (A) Contour-plot of a narrow-band image centred on the
continuum adjacent to the [OIII] emission line (5204-5246\AA) extracted from
the residual data cube, together with a greyscale of the residual from the 2D
modeling of the F555W-band image (Fig. 3D). (B) Contour-plot of a
narrow-band image centred on the [OIII]$\lambda$5007 emission line
(5170-5200\AA) extracted from the residual data cube, together with a
greyscale of the F555W-F547M image shown in Fig. 3D. The different
continuum dominated (left) and emission-line dominated (right) substructures
are labelled with their corresponding names. 

\plotone{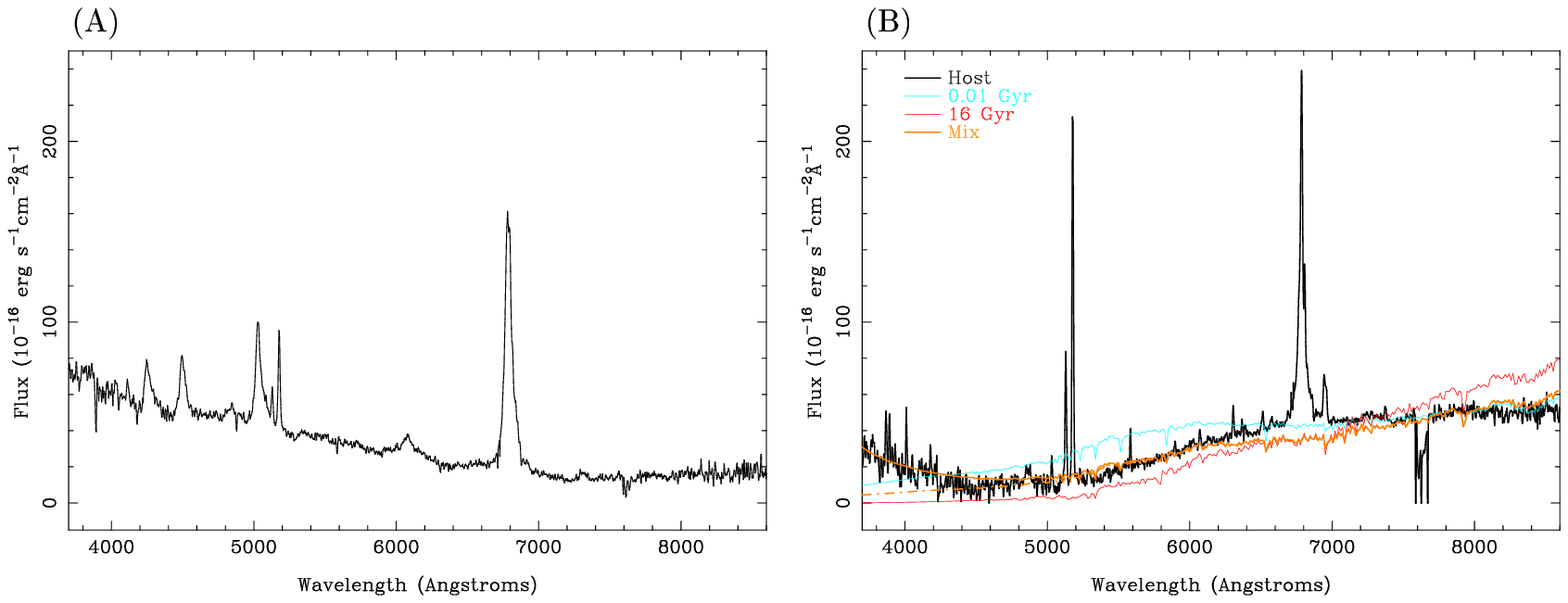}
{\bf Figure 8:} (A) Spectrum of the nucleus of 3C~120, once decontaminated
from the contribution of the host galaxy. (B) Spectrum of the host galaxy
(black line) together with the spectrum of synthetic models for single stellar
populations of 16 Gyr (red line), 0.01 Gyr (blue line), and a mix of both
(orange line), adding (solid line) and not adding (dashed line) a power-law
continuum to simulate scattered light from the nucleus.

\plotone{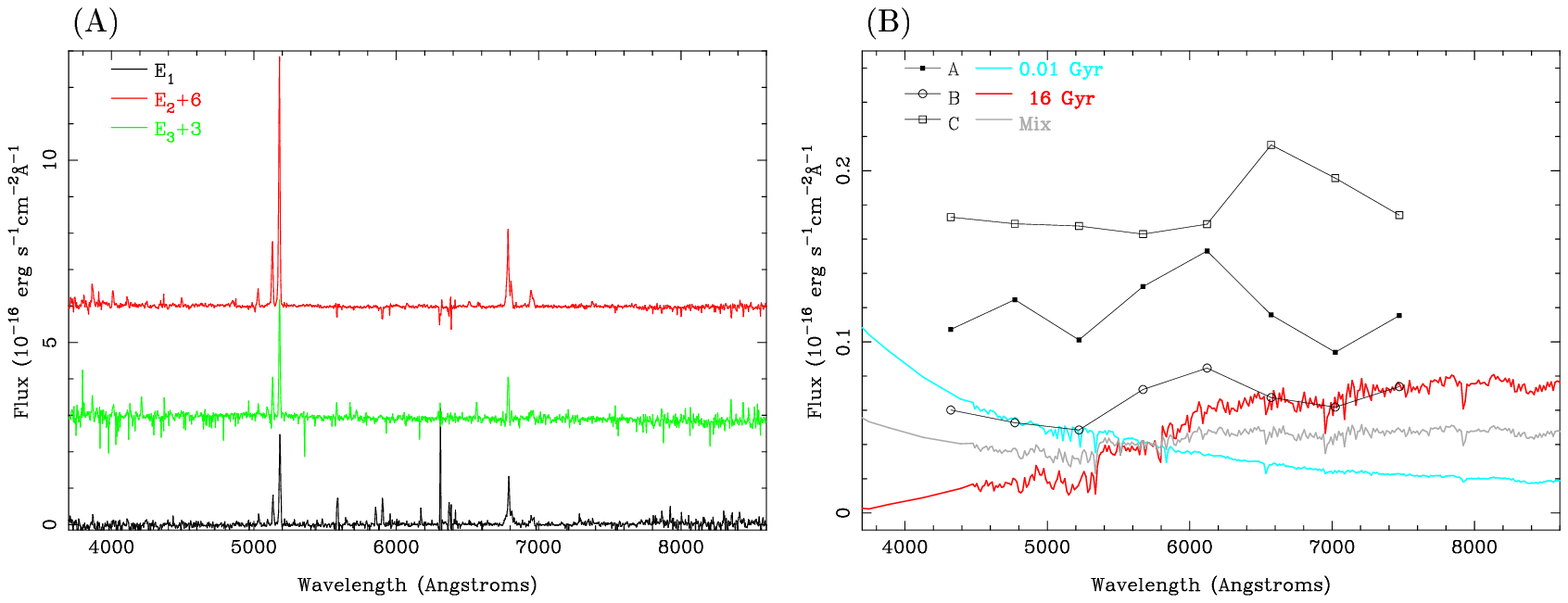}
{\bf Figure 9:} (A) Spectra of the different EELRs detected in the residual
data cube. (B) Spectra of the different continuum-dominated structures
detected in the residual data cube (points) and a median filtered version of
these data (solid lines, solid squares), with a 300\AA\ width. 

\epsscale{1}
\plotone{f10.ps}
{\bf Figure 10:} [OIII]$\lambda$5007/H$\beta$ vs. [NII]$\lambda$6583/H$\alpha$ intensity ratios
for the different components of 3C~120. Solid curve divides AGNs from
H$_\mathrm{II}$ region-like objects.

\epsscale{0.8}
\plotone{f11.ps}

{\bf Figure 11:} Ionized gas velocity fields of 3C~120
derived from the Gaussian fit to: (A) [OIII], isovelocity lines extend
from 10200 to 10600 in steps of 50 \kms; (B) H$\alpha$ where
isovelocity lines extend from 10000 to 10600 in steps of 50 km s${-1}$. (C) As (A) with the [OIII] intensity of E$_1$, E$_2$,
E$_3$ in black contours. (D) The velocity distribution of ionized gas
structures (E$_1$, E$_2$, E$_3$). Isovelocity lines extend from 9600
to 10300 in steps of 100 \kms (E) The simplest rotational model
obtained considering an inclination of 40$^{\circ}$ and PA=-40$^{\circ}$. (F)
The residual velocity map after the subtraction of the rotational
model (E) from the ionized gas velocity field (A). Isovelocity lines
corresponds to kinematics of structures E$_1$, E$_2$, E$_3$ as in
(D). 
\epsscale{1}

\plotone{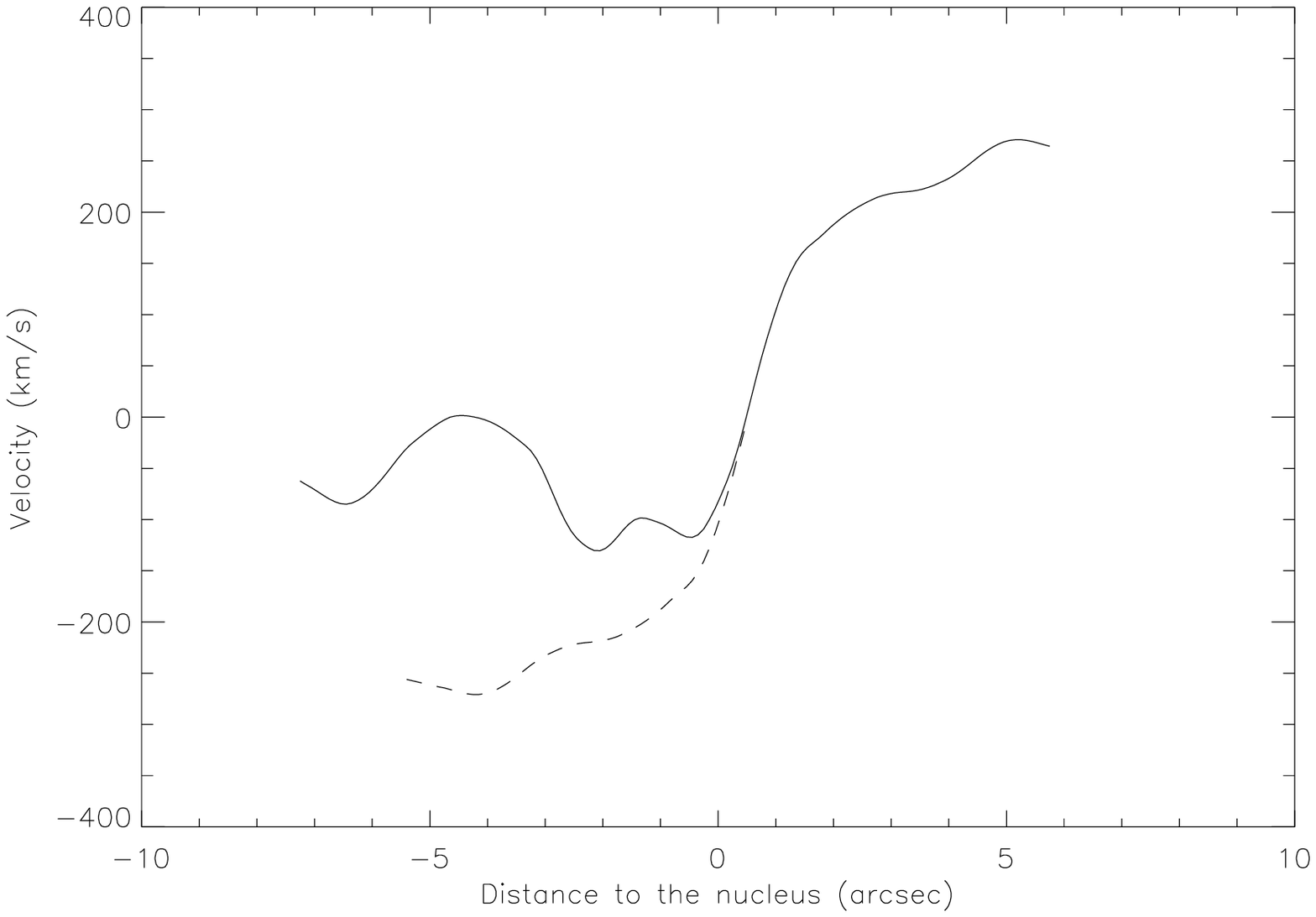}
{\bf Figure 12:}  The solid line shows the H$\alpha$ velocity distribution of
3C~120 along a position angle of $-40^{\circ}$ from south-east (left) to
north-west (right). The perturbation associated with E$_2$ is seen as a peak at
approximately $-5\arcsec$. The dashed-line shows the distribution obtained
assuming that the north-west portion of the velocity distribution describes the behavior of the rotating component of the galaxy, and modeling the south-east
by a symmetrical distribution.

\plotone{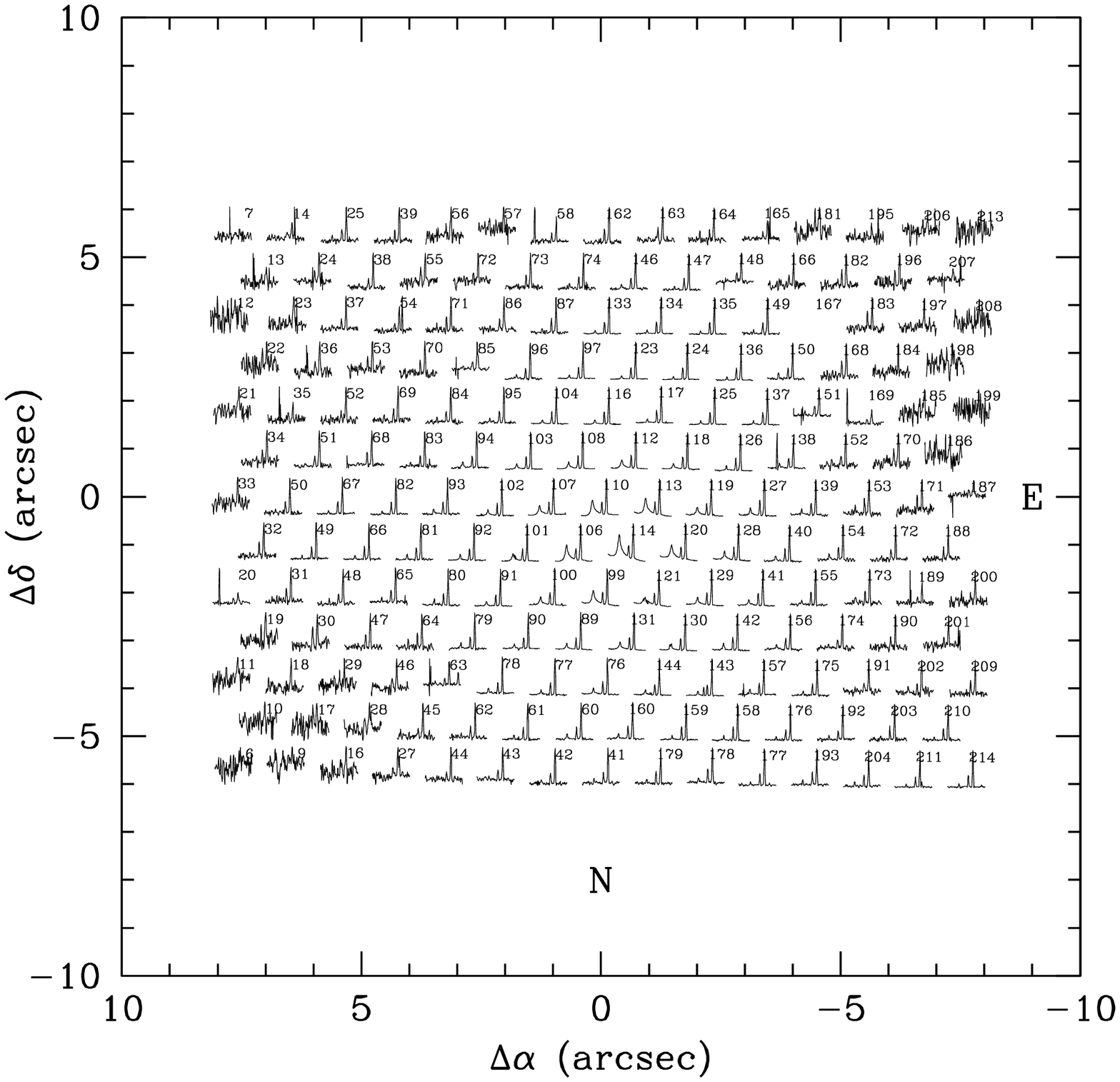}
{\bf Figure 13:}  Two-dimensional spectroscopic diagram of
H$\beta$+[OIII]$\lambda\lambda4959,5007$ emission line profiles. Plotted spectral range: 4910-5300 \AA . 

\plotone{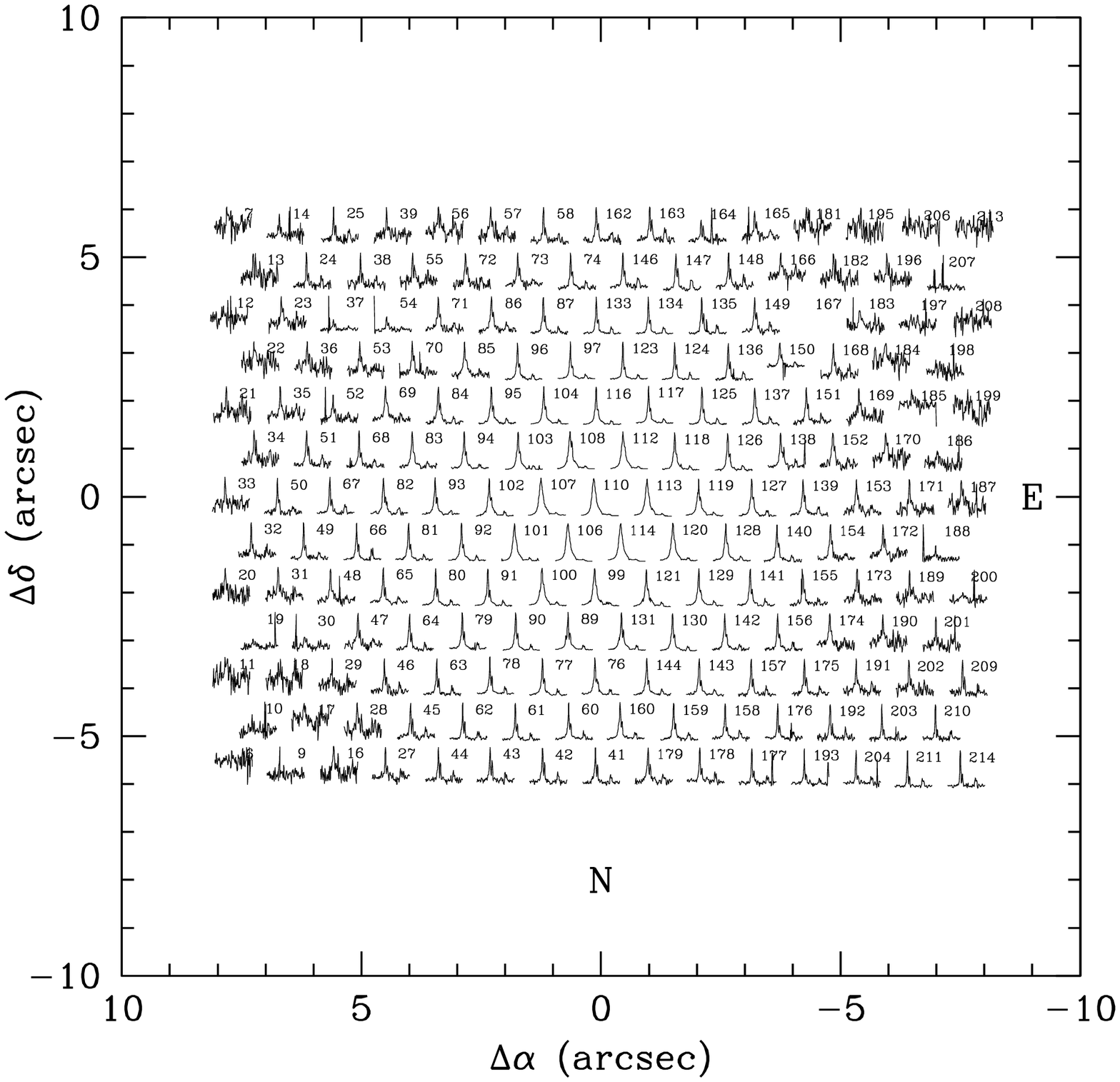}
{\bf Figure 14:} Two-dimensional distribution of emission line profiles of\\ H$\alpha$+[NII]$\lambda\lambda6548,6584$+[SII]$\lambda\lambda6716,6730$. The spectral range plotted is 6650-7050 \AA . 

\epsscale{0.95}
\plotone{f15.ps}
{\bf Figure 15:}   Centroid coordinates, {\it x} and {\it y}, in pixels
  (0$\farcs$3/pixel), as a function of the wavelength as derived from the
  first run of the 3D modeling. The solid-lines show the result of the
  fitting with a polynomial function. 
\epsscale{1}

\plotone{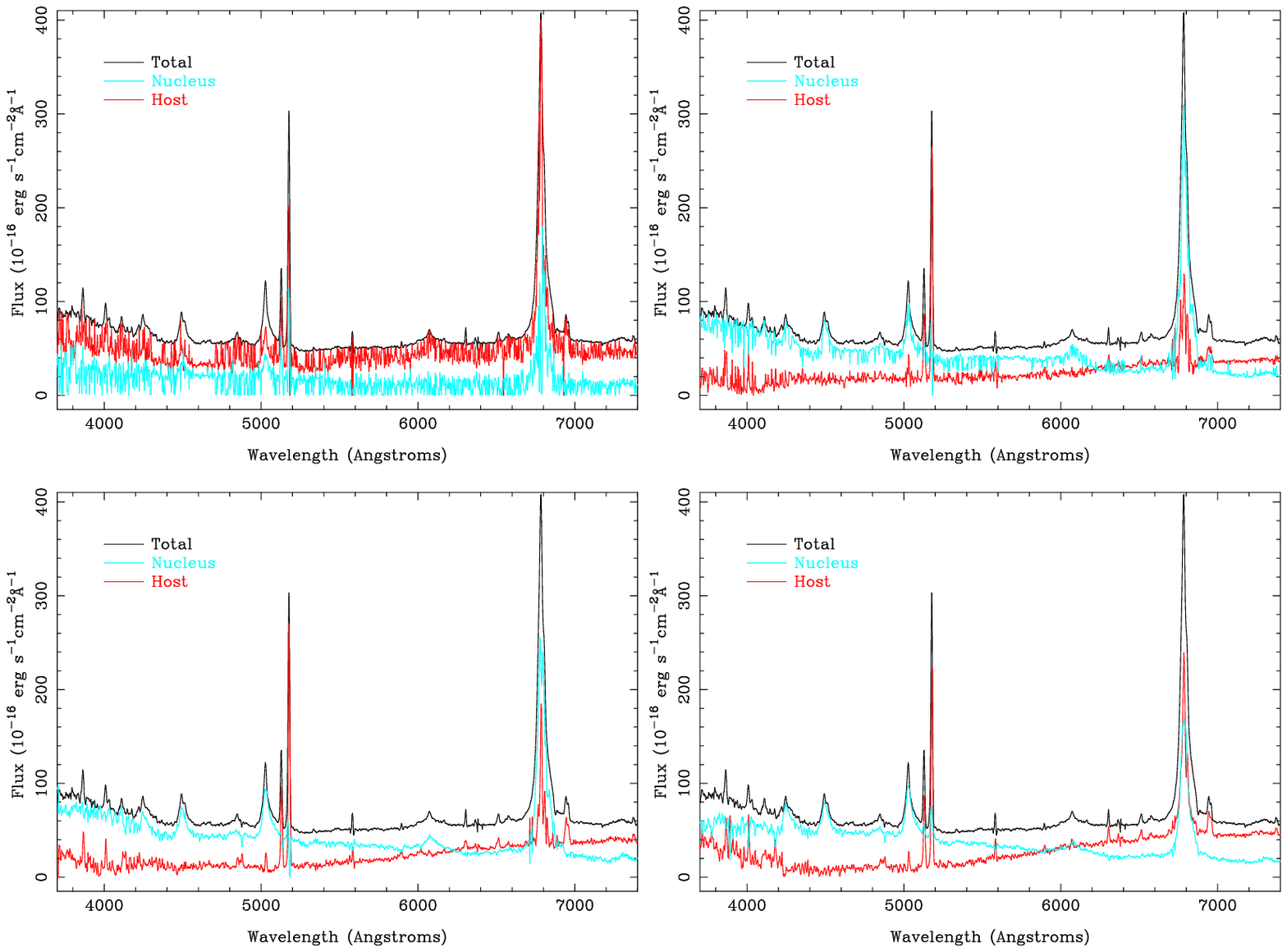}
 {\bf Figure 16:}  Top-left panel: Decoupled spectra of the nucleus and the host of 3C~120,
  together with the total spectrum of the object, derived from the 3D
  modeling when all the parameters are fitted freely. Top-right panel:
  Similar spectra derived from the modeling when the centroid is fixed to the
  result of the polynomial function fitting shown in Fig. 15.
  Bottom-left panel: Similar spectra derived when in addition to the
  centroids, the ellipticity, position angle and effective radius of the host
  galaxy are fix to the result of a polynomial function fitting as a function
  of the wavelength. Bottom-right panel: Final decoupled spectra obtained when
  the structural parameters of the host galaxy are fixed to the values derived
  from the 2D modeling of the HST images.

\plotone{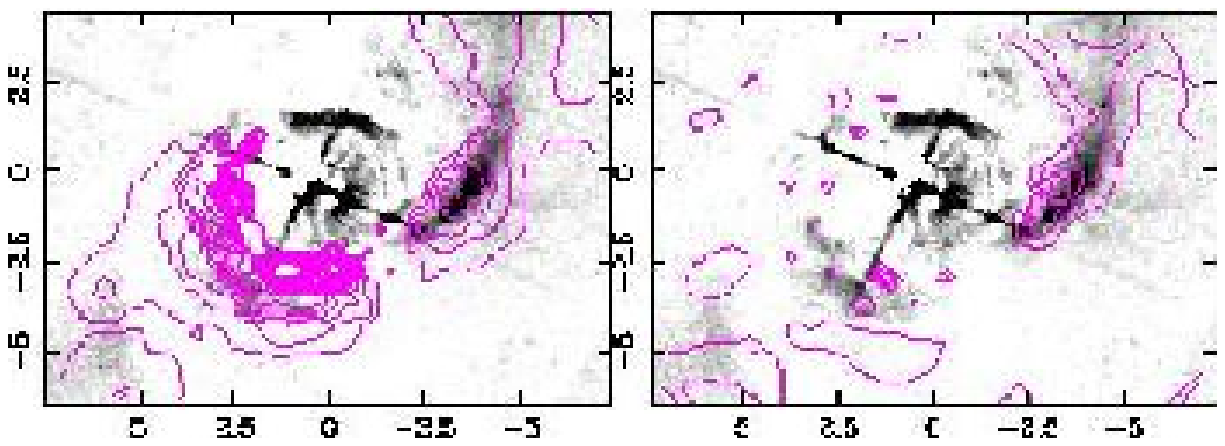}
 {\bf Figure 17:}  Left panel: Contour-plot of a narrow-band image centred on the continuum
  adjacent to the [OIII] emission line (5204-5246\AA) from the residual
  datacube obtained by the 3D modeling. Right panel:
  Contour-plot of a narrow-band image centred on the continuum adjacent to the
  [OIII] emission line (5204-5246\AA) from the residual datacube obtained by
  the 3D surface brightness analysis. In both panels the grayscale shows the
  residual from the 2D modeling of the F555W-band image
  (Fig. 3D)

\plotone{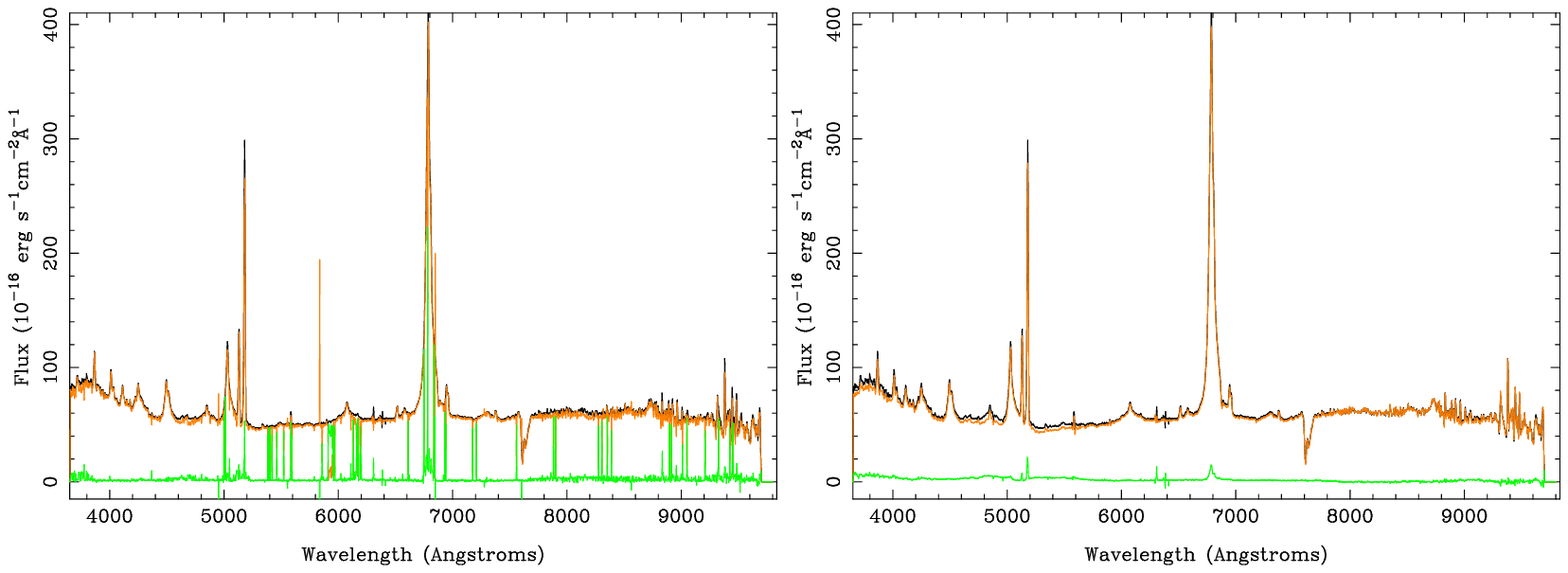}
 {\bf Figure 18:}  Left panel: Integrated spectrum of 3C~120 (black-line), together with
  the spectrum of the model (orange) and the residuals (green),
  obtained using the 3D surface brightness analysis. Right panel: Similar
  spectra than the ones presented in the left panel obtained once  the
  structural parameters are fixed to the result of a polynomial function fitting as a
  function of the wavelength.

\end{document}